%% file: ms.tex
\documentclass[iop,revtex4]{emulateapj}
\usepackage{lscape}
\usepackage{multirow}
\usepackage{natbib}
\newcommand{\ee}[1]{\mbox{${} \times 10^{#1}$}}% scientific number format
\newcommand{\eten}[1]{\mbox{$10^{#1}$}}% power of ten
\newcommand{\lbol}{\mbox{$L_{bol}$}} % bolometric luminosity
\newcommand{\lsmm}{\mbox{$L_{smm}$}} % luminosity longward of 350 mic.
\newcommand{\lint}{\mbox{$L_{int}$}} % internal luminosity
\newcommand{\lacc}{\mbox{$L_{acc}$}} % accretion luminosity
\newcommand{\tbol}{\mbox{$T_{bol}$}} % bolometric temperature
\newcommand{\lbolsmm}{\mbox{$L_{bol}/L_{smm}$}} % bol-to-smm luminosity
\newcommand{\lsmmbol}{\mbox{$L_{smm}/L_{bol}$}} % smm-to-bol luminosity
\newcommand{\tbolprime}{\mbox{$T^{\prime}_{bol}$}} % Tbol after ext. corr.
\newcommand{\lbolprime}{\mbox{$L^{\prime}_{bol}$}} % Lbol after ext. corr.
\newcommand{\aprime}{\mbox{$\alpha^{\prime}$}} % alpha after ext. corr.
\newcommand{\av}{\mbox{$A_V$}} % Visual Extinction
\newcommand{\ta}{{$T_A^*$}}
\newcommand{\tmb}{\mbox{$T_{\rm mb}$}}
\newcommand{\etamb}{$\eta_{\rm mb}$}
\newcommand{\andre}{Andr\'{e}}
\newcommand{\merin}{Mer\'{i}n}
\newcommand{\kovacs}{Kov\'{a}cs}
\newcommand{\comeron}{Comer\'{o}n}
\newcommand{\jorgensen}{J{\o}rgensen}
\newcommand{\alcala}{Alcal{\'a}}
\newcommand{\csotau}{$\tau_{225\ GHz}$}

\newcommand{\degree}{\mbox{$^{\circ}$}}
\newcommand{\am}{\mbox{\arcmin}}
\newcommand{\as}{\mbox{\arcsec}}
\newcommand{\cms}{\mbox{cm s$^{-1}$}}% cm/s
\newcommand{\kms}{\mbox{km s$^{-1}$}}% km/s
\newcommand{\jybeam}{\mbox{Jy beam$^{-1}$}}% Jy/beam
\newcommand{\mjybeam}{\mbox{mJy beam$^{-1}$}}% Jy/beam
\newcommand{\mjysr}{\mbox{MJy sr$^{-1}$}}% Jy/beam
\newcommand\cmv{\mbox{cm$^{-3}$}}
\newcommand\cmc{\mbox{cm$^{-2}$}}
\newcommand\cmdv{\mbox{cm$^{-2}$ (\kms)$^{-1}$}}
\newcommand{\um}{$\mu$m}
\newcommand{\lsun}{\mbox{L$_\odot$}}% Lsun
\newcommand{\msun}{\mbox{M$_\odot$}}% Msun
\newcommand{\rsun}{\mbox{M$_\odot$}}% Msun

\newcommand{\hh}{\mbox{{\rm H}$_2$}}
\newcommand{\form}{H$_2$CO}
\newcommand{\water}{H$_2$O}
\newcommand{\ammonia}{\mbox{{\rm NH}$_3$}}
\newcommand{\co}{$^{12}$CO}
\newcommand{\coo}{$^{13}$CO}
\newcommand{\cooo}{C$^{18}$O}
\newcommand{\coooo}{C$^{17}$O}
\newcommand{\hcop}{HCO$^+$}
\newcommand{\hcoop}{H$^{13}$CO$^+$}
\newcommand{\dcop}{DCO$^+$}
\newcommand{\nthp}{N$_2$H$^+$}
\newcommand{\ntdp}{N$_2$D$^+$}
\newcommand{\cojone}{$^{12}$CO (1--0)}
\newcommand{\cojtwo}{$^{12}$CO (2--1)}
\newcommand{\cojthree}{$^{12}$CO (3--2)}
\newcommand{\cojsix}{$^{12}$CO (6--5)}
\newcommand{\cojten}{$^{12}$CO (10--9)}
\newcommand{\coojtwo}{$^{13}$CO (2--1)}
\newcommand{\coojthree}{$^{13}$CO (3--2)}
\newcommand{\cooojtwo}{C$^{18}$O (2--1)}
\newcommand{\hcopjthree}{HCO$^+$ (3--2)}

\newcommand{\cjaa}{ChJAA}
\newcommand{\caa}{ChA\&A}

\newcommand{\spitzer}{\emph{Spitzer}}
\newcommand{\herschel}{\emph{Herschel}}
\newcommand{\iras}{\emph{IRAS}}
\newcommand{\iso}{\emph{ISO}}

\slugcomment{Accepted for publication in ApJS}

\begin{document}
%%%%%%%%%%%%%%%%%% title %%%%%%%%%%%%%%%%%%%%%%%%%%%%%%%%%%%%%%%%
\title {Young Stellar Objects in the Gould Belt}

\author{
Michael M.~Dunham\altaffilmark{1,2}, 
Lori E.~Allen\altaffilmark{3}, 
Neal J.~Evans II\altaffilmark{4}, 
Hannah Broekhoven-Fiene\altaffilmark{5}, 
Lucas A.~Cieza\altaffilmark{6,7}, 
James Di~Francesco\altaffilmark{8}, 
Robert A.~Gutermuth\altaffilmark{9}, 
Paul M.~Harvey\altaffilmark{4}, 
Jennifer Hatchell\altaffilmark{10}, 
Amanda Heiderman\altaffilmark{11,12}, 
Tracy L.~Huard\altaffilmark{13,14}, 
Doug Johnstone\altaffilmark{8,15}, 
Jason M.~Kirk\altaffilmark{16}, 
Brenda C.~Matthews\altaffilmark{8}, 
Jennifer F.~Miller\altaffilmark{17}, 
Dawn E.~Peterson\altaffilmark{18}, 
and 
Kaisa E.~Young\altaffilmark{19}
}

\altaffiltext{1}{Harvard-Smithsonian Center for Astrophysics, 60 Garden Street, MS 78, Cambridge, MA 02138, USA}

\altaffiltext{2}{mdunham@cfa.harvard.edu}

\altaffiltext{3}{National Optical Astronomy Observatories, Tucson, AZ, USA}

\altaffiltext{4}{Department of Astronomy, The University of Texas at Austin, 2515 Speedway, Stop C1400, Austin, TX 78712-1205, USA}

\altaffiltext{5}{Department of Physics \& Astronomy, University of Victoria, Victoria, BC, V8W 3P6, Canada}

\altaffiltext{6}{N\'{u}cleo de Astronom\'{i}ía de la Facultad de Ingenier\'{i}ía, Universidad Diego Portales, Av.~Ej\'{e}rcito 441, Santiago, Chile}

\altaffiltext{7}{Millenium Nucleus ``Protoplanetary Disks in ALMA Early Science,'' Chile}

\altaffiltext{8}{National Research Council of Canada, Herzberg Astronomy \& Astrophysics Programs, 5071 West Saanich Road, Victoria, BC, Canada V9E 2E7}

\altaffiltext{9}{Department of Astronomy, University of Massachusetts, Amherst, MA 01003, USA}

\altaffiltext{10}{Physics and Astronomy, University of Exeter, Stocker Road, Exeter EX4 4QL, UK}

\altaffiltext{11}{Department of Astronomy, University of Virginia, P.O. Box 400325, Charlottesville, VA 22904, USA}

\altaffiltext{12}{National Radio Astronomy Observatory, 520 Edgemont Road, Charlottesville, VA 22903, USA}

\altaffiltext{13}{Department of Astronomy, University of Maryland, College Park, MD 20742, USA}

\altaffiltext{14}{Columbia Astrophysics Laboratory, Columbia University, New York, NY 10027, USA}

\altaffiltext{15}{Department of Physics and Astronomy, University of Victoria, Victoria, BC V8P 1A1, Canada}

\altaffiltext{16}{Jeremiah Horrocks Institute, University of Central Lancashire, Preston, PR1 2HE, United Kingdom}

\altaffiltext{17}{Gemini Observatory, 670 N. A'ohoku Pl, Hilo, HI 96720, USA}

\altaffiltext{18}{Space Science Institute, 4750 Walnut Street, Suite 205, Boulder, CO 80301, USA}

\altaffiltext{19}{Department of Physical Sciences, Nicholls State University, PO Box 2022, Thibodaux, LA 70310}

\begin{abstract}
We present the full catalog of Young Stellar Objects (YSOs) identified in the 
18 molecular clouds surveyed by the {\it Spitzer Space Telescope} ``cores to 
disks'' (c2d) and ``Gould Belt'' (GB) Legacy surveys.  Using standard 
techniques developed by the c2d project, we identify 3239 candidate YSOs in the 
18 clouds, 2966 of which survive visual inspection and form our final catalog 
of YSOs in the Gould Belt.  We compile extinction 
corrected SEDs for all 2966 YSOs and calculate and tabulate the infrared 
spectral index, bolometric luminosity, and bolometric temperature for each 
object.  We find that 326 (11\%), 210 (7\%), 1248 (42\%), and 1182 (40\%) 
are classified as Class 0+I, Flat-spectrum, Class II, and Class III, 
respectively, and show that the Class III sample suffers from 
an overall contamination rate by background AGB stars between 25\% and 90\%.  
Adopting standard assumptions, we derive durations of $0.40-0.78$ Myr for 
Class 0+I YSOs and $0.26-0.50$ Myr for Flat-spectrum YSOs, where the ranges 
encompass uncertainties in the adopted assumptions.  Including information 
from (sub)millimeter wavelengths, one-third of the Class 0+I sample is 
classified as Class 0, leading to durations of $0.13-0.26$ Myr (Class 0) and 
$0.27-0.52$ Myr (Class I).  We revisit infrared color-color diagrams used in 
the literature to classify YSOs and propose minor revisions to classification 
boundaries in these diagrams.  Finally, we show that the bolometric 
temperature is a poor discriminator between Class II and Class III YSOs.
\end{abstract}


%%%%%%%%%%%%%%%%%%%%%%%%%%%%%%%%%%%%%%%%%%%%%%%%%%%%%%%%%%%%%

\section{Introduction}\label{sec_intro}

Most of the current star formation within 500 pc of the Sun occurs in the 
Gould Belt, a ring of nearby O-type stars inclined approximately 20\degree\ 
with respect to the Galactic Plane \citep{herschel1847:gould,gould1879:gould}.  
The Sun is located interior to the Gould Belt, which has a radius of about 350 
pc \citep[e.g.,][]{clube1967:gould,stothers1974:gould,comeron1992:gould,dezeeuw1999:gould,poppel2001:gould}.  
All of the nearby, well-studied giant molecular clouds (e.g., Ophiuchus, 
Taurus, Perseus, Serpens, Orion, etc.) are located within this ring 
\citep[see Figure 1 of][for a schematic diagram of the Gould Belt in galactic coordinates]{wardthompson2007:gould_scuba2}.

Due to the high visual extinction toward giant molecular clouds, a complete 
census of star formation in these clouds requires infrared observations.  The 
{\it Infrared Astronomical Satellite (IRAS)} provided the first such unbiased 
survey of the Gould Belt (and the rest of the sky), but was limited in both 
sensitivity and spatial resolution.  For young stellar objects (YSOs) still 
in the embedded phases of evolution, \iras\ was only sensitive to objects 
with luminosities above 0.1 $(d/140 {\rm pc})^2$ \lsun\ 
\citep{myers1987:iras}. For more evolved YSOs, \iras\ was even less sensitive.  
The spatial resolution provided by \iras\ ranged from 0.5\am\ at 
12 \um\ ($0.02-0.07$ pc at distances of $150-500$ AU) to 2\am\ at 100 \um\ 
($0.09-0.29$ pc at distances of $150-500$ AU), insufficient to resolve 
individual stars forming in clustered regions of molecular clouds.  
Ground-based near-infrared surveys 
\citep[e.g., the Two Micron All Sky Survey (2MASS);][]{skrutskie2006:2mass} 
provide much higher spatial resolution but are unable to detect the most 
deeply embedded, heavily extincted, and/or lowest luminosity YSOs.  
Ground-based mid-infrared studies are generally limited by the atmosphere to 
only the brightest and most luminous objects.  Finally, selected regions 
of the Gould Belt were mapped in the mid-infrared by the {\it Infrared Space 
Observatory (ISO)}. While {\it ISO} surveys identified many new deeply embedded 
and low luminosity YSOs, they only focused on small portions of molecular 
clouds \citep[e.g.,][]{bontemps2001:iso,persi2003:iso,kaas2004:iso}.

The launch of the {\it Spitzer Space Telescope} \citep{werner2004:spitzer} 
in 2003 opened new avenues for infrared surveys of giant molecular clouds.  
With its combination of high sensitivity and large format detector arrays 
operating at $3.6-160$ \um, {\it Spitzer} was the first mid-infrared facility 
capable of mapping entire molecular clouds within reasonable amounts of time.  
The {\it Spitzer} Legacy project ``“From Molecular Cores to Planet-forming 
Disks'' (Cores to Disks, or c2d) took advantage of these features by mapping 
seven large, nearby Gould Belt molecular clouds and nearly 100 isolated dense 
cores.  An overview of the survey is given by \citet{evans2003:c2d} and a 
summary of results for the large clouds is given by \citet{evans2009:c2d}.  
The c2d survey identified 
over 1000 YSOs, many of which were newly identified, and demonstrated 
{\it Spitzer's} ability to detect YSOs that are too deeply embedded, 
too low luminosity, and/or too low mass to be detected by previous facilities.

The {\it Spitzer} Gould Belt (GB) Legacy survey followed up the c2d survey 
by mapping an additional 11 Gould Belt clouds using the same 
observation and data reduction strategies as c2d.  Combined, the two 
surveys targeted nearly every major site of star formation within 500 pc of the 
Sun.  The only two exceptions are the Taurus and Orion Molecular Clouds, 
both of which are so large they were the subject of their own, dedicated 
surveys by different teams \citep{rebull2010:taurus,megeath2012:orion}.  
Adding the results of the GB survey to those obtained by c2d nearly 
triples the number of YSOs identified in the Gould Belt.  
A preliminary attempt to combine the results from the c2d, GB, Taurus, and 
Orion surveys was presented by \citet{dunham2014:ppvi}, but a full combination, 
including reconciliation of different methods for observing, 
data reduction, source extraction and classification, is left for future work 
that combines the expertise of the respective surveys.  Here we only consider 
the c2d and GB surveys.

While the full list of YSOs identified by \spitzer\ in the c2d and GB surveys 
has been used in several studies, including those on clustering of young 
stars \citep{bressert2010:clustering}, star formation rates and thresholds 
\citep{heiderman2010:threshold}, the luminosities of protostars 
\citep{dunham2013:luminosities}, relationships between gas and star formation 
\citep{evans2014:sfrelations}, and the physical nature of empirically 
defined classes of young stars \citep{heiderman2015:misfits}, 
this list itself has not yet appeared in publication.  To rectify this, we 
present in this paper the full list of YSOs in the combined 
c2d and GB clouds.  For the c2d clouds, this list updates and replaces the 
previous version given by \citet{evans2009:c2d}, including an analysis 
of contamination by background AGB stars, updated extinction corrections, 
and revised spectral energy distributions (SEDs).  Although we use this YSO 
list to investigate classification of young stars and the durations of these 
classes, the primary purpose of this work is to present a full catalog of 
{\it Spitzer} c2d+GB YSOs.  While useful on its own, this YSO list will reach 
its full potential in combination with results from the far-infrared and 
submillimeter surveys of the Gould Belt by the {\it Herschel Space 
Observatory} and SCUBA-2 on the James Clerk Maxwell Telescope, once they 
become fully available 
\citep{wardthompson2007:gould_scuba2,andre2010:gb,harvey2013:auriga}.  
By providing the full {\it Spitzer} YSO list, we enable such comparisons.

The organization of this paper is as follows.  We describe our method 
in \S \ref{sec_method}, including an overview of the c2d and GB surveys 
(\S \ref{sec_method_surveys}), the procedure for identifying YSOs 
(\S \ref{sec_method_yso}), residual contamination in our YSO sample by 
background AGB stars (\S \ref{sec_method_contamination}),
 and the construction of full SEDs and corrections 
for extinction (\S \ref{sec_method_seds}).  We present our results in \S 
\ref{sec_results}, including the full list of YSOs (\S \ref{sec_list}) and 
a discussion of classification of young stars 
(\S \ref{sec_classification}$-$ \ref{sec_cc}).  We discuss timescales for 
the classes of young stars in \S \ref{sec_lifetimes}.  Finally, we summarize 
our results in \S \ref{sec_summary}.

\section{Method}\label{sec_method}

A concise summary of these surveys, the adopted procedure for identifying young 
stellar objects, and the methods used to construct SEDs and to correct the 
photometry for foreground extinction are given in Section 2 of 
\citet{dunham2013:luminosities}.  Here we repeat the key information and refer 
the reader to Dunham et al.~for more details.

\subsection{Overview of the Surveys}\label{sec_method_surveys}

The \emph{Spitzer} c2d survey (PI: N.~J.~Evans) imaged seven large, nearby 
molecular clouds in the Gould Belt and approximately 100 isolated dense 
molecular cores.  \citet{evans2003:c2d} review both the primary science 
objectives of c2d and its basic observation plan, and 
\citet{evans2009:c2d} summarize the results in the large clouds.  
The \emph{Spitzer} GB survey (PI: L.~E.~Allen) is a follow-up program that 
imaged an additional 11 molecular clouds in the Gould Belt, completing most of 
the remaining clouds in the Gould Belt except for the Taurus and Orion 
molecular clouds (see \S \ref{sec_intro}).  Both surveys imaged their 
respective targets at 3.6--8.0 \um\ with the \emph{Spitzer} Infrared Array 
Camera \citep[IRAC;][]{fazio2004:irac}, and at 24--160 \um\ images with the 
Multiband Imaging Photometer \citep[MIPS;][]{rieke2004:mips}.  A data pipeline 
developed by c2d was applied to the observations obtained by both programs.  
This pipeline includes data reduction, source extraction, and band-merging 
the results from each wavelength to produce final images and source catalogs, 
and has been described in detail elsewhere 
\citep{harvey2006:serpens,evans2007:deldoc}.

For each cloud targeted by c2d or GB, Table \ref{tab_clouds} lists the parent 
survey, the assumed distance and reference for the distance, the total area 
mapped, the total cloud gas mass, and references to in-depth studies of each 
cloud based on the c2d or GB data, when available.  The \spitzer\ 
Astronomical Observation Requests (AORs) for each cloud can be found in 
these references to individual cloud studies; Appendix A lists the AORs for 
the clouds without published data references.  The isolated cores 
surveyed by c2d are not considered here except for a few cases where 
the cores are actually parts of larger cloud complexes and included in the 
results for those clouds \citep[namely Cepheus and Ophiuchus North;][]{kirk2009:cepheus,hatchell2012:ophnorth}.  Note that the distances 
to these clouds are not all well determined, and some cloud distances 
are still under debate.  One such example is the uncertainty over the distance 
to Aquila, with two distances (260 pc and 429 pc) commonly adopted in the 
literature \citep[see, e.g.,][for details]{harvey2006:serpens,gutermuth2008:serpsouth,dzib2011:dserpens,maury2011:aquila}.  For this reason we do not list 
formal distance uncertainties and instead refer the reader to the 
references given in Table \ref{tab_clouds} for full discussions of the 
distances to each cloud.

The total cloud areas mapped and total cloud gas masses 
contained within these areas are taken from \citet{heiderman2010:threshold}, 
with the masses determined from extinction maps (see Heiderman et al.~for 
details).  For most clouds, the mapped areas were chosen 
to map the entire cloud above a minimum visual extinction in the range 
2 $<$ A$_{\rm V}$ $<$ 3, with the exact value for each cloud depending on 
the total area that could be mapped within the allocated time.  For Serpens 
and Aquila, only the cloud areas above A$_{\rm V} = 6$ were mapped due to time 
constraints and difficulty in separating the cloud from the general 
extinction close to the Galactic midplane.

\subsection{Identification of Young Stellar Objects}\label{sec_method_yso}

The data reduction pipeline used by both the c2d and GB surveys creates 
band-merged source catalogs where the sources extracted at each {\it Spitzer} 
wavelength are matched with each other and with the 2MASS catalog, creating a 
final catalog of $1.25-70$ \um\ photometry for each cloud.  
Candidate YSOs are identified using 
a method developed by \citet{harvey2007:serpens} and summarized in all of 
the publications focused on individual clouds that are listed in Table 
\ref{tab_clouds}.  In this method, the {\it Spitzer} SWIRE Legacy survey 
of the ELAIS N1 extragalactic field \citep{lonsdale2003:swire} is processed 
to simulate the same sensitivity and extinction distribution of the c2d and 
GB clouds, and then used to locate background galaxies observed through the 
target clouds in various infrared color-magnitude diagrams.  
Using these results to define the expected colors 
and magnitudes of background galaxies, each source extracted in the cloud 
maps with infrared colors that can not be explained as an extincted background 
star is assigned an unnormalized ``probability'' of being a galaxy or YSO 
based on its position in each color-magnitude diagram, its morphology in the 
two shortest \emph{Spitzer} IRAC bands (3.6 and 4.5 \um), and its 24 and 70 
\um\ flux densities.  Since it is unnormalized, this quantity is not a true 
probability but a relative parameter expressing the likelihood of being a 
background galaxy \citep[higher values indicate higher likelihood of being 
galaxies; see][for details]{harvey2007:serpens}.  
The final boundary between candidate YSO 
and candidate galaxy in this quantity is set to provide a nearly complete 
elimination of SWIRE sources, likely at the cost of excluding YSOs with 
overlapping colors and magnitudes \citep[e.g.,][]{hsieh2013:ysos}.  

One potential drawback to the c2d+GB method of identifying YSOs is that it 
requires detections in all four IRAC bands ($3.6-8.0$ \um) as well as the first 
MIPS band (24 \um).  As the 24 \um\ MIPS band has lower resolution and 
sensitivity, some of the faintest and/or most clustered YSOs may be missed 
\citep[e.g.,][Gutermuth et al., in preparation.]{gutermuth2009:ysos}.  
We note that several alternative classification methods for separating 
background galaxies and YSOs have been presented in the literature 
\citep[e.g.,][]{gutermuth2009:ysos,rebull2010:taurus,kryukova2012:luminosities,hsieh2013:ysos}.  While they are all based on a similar philosophy of 
separating sources in various colors and magnitudes, they differ in their 
detailed implementations of this philosophy.  A full comparison of these 
methods and ultimate consensus on best practices for extracting YSOs from 
infrared surveys of star-forming clouds is left for future work.

Applying these selection criteria results in the identification of 3239 
candidate YSOs in the 18 c2d and GB clouds.  We note here that we only consider 
sources that are located in the overlap regions imaged by both the IRAC and 
MIPS instruments, as these are the sources that can be most reliably classified 
using the method summarized above.  For technical reasons, the MIPS 
observations often covered much larger areas (see the references listed in 
Table \ref{tab_clouds} for more details on the observations of each individual 
cloud).  Several of the individual cloud studies listed in Table 
\ref{tab_clouds} also present candidate YSO lists in the MIPS-only regions 
using alternative identification and classification techniques, but such 
sources are not considered here.  We visually inspected each of the 3239 
candidate YSOs to remove resolved galaxies and image artifacts that 
are sometimes misidentified \citep[see][for details]{evans2009:c2d}.  
Additionally, 
a few known YSOs that were saturated or undetected at one or more wavelengths 
were missing from the initial list and were added by hand.  In the end, we were 
left with a final list of 2966 YSOs, and since all have passed visual 
inspection, we follow \citet{evans2009:c2d} and drop the prefix ``candidate'' 
at this point.  

Finally, we note that initial results from \herschel\ surveys of the c2d and GB 
clouds suggest that a modest number of the most deeply embedded protostars are 
undetected by \spitzer, increasing the number of Class 0 
protostars\footnote{Class 0 protostars are the youngest class of YSOs; see \S 
\ref{sec_list} for the formal definition of this class of objects.} by 15\% -- 
30\% \citep[e.g.,][]{harvey2013:auriga,stutz2013:pbrs,sadavoy2014:perseus}.  
Revisions to the YSO lists presented here will thus be necessary once complete 
\herschel\ results are available for each cloud.

\subsection{Background Giant Star Contamination}\label{sec_method_contamination}

\begin{figure*}
\epsscale{1.2}
\plotone{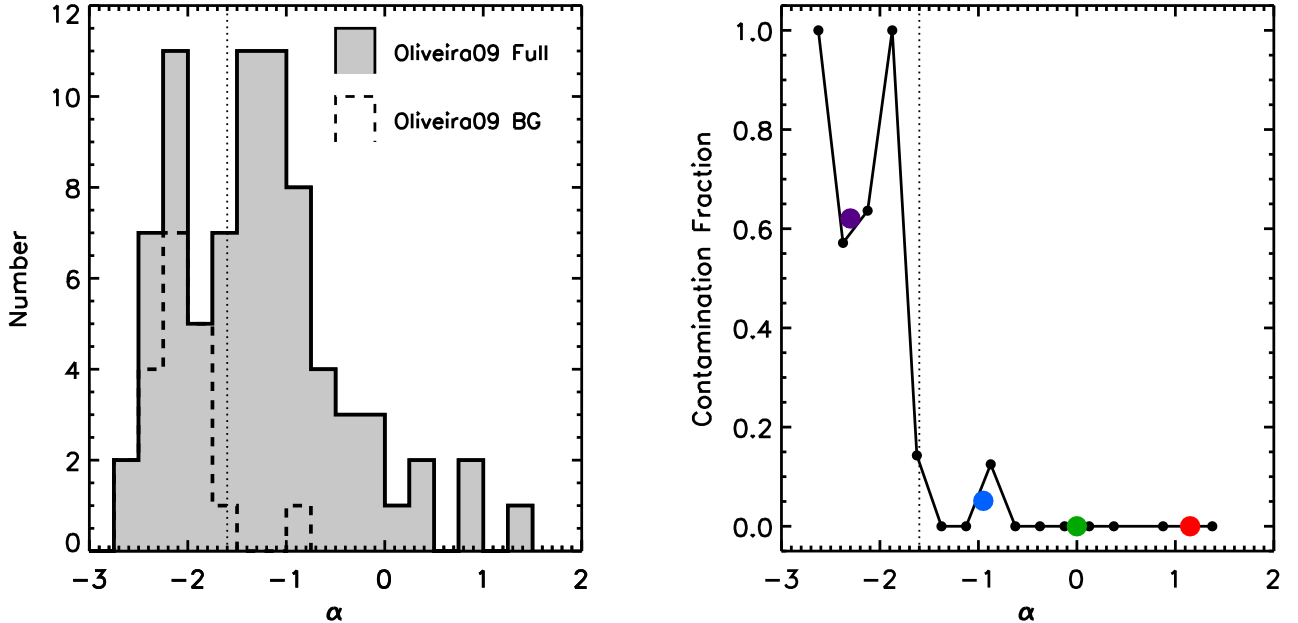}
\caption{\label{fig_contamination}{\it Left:} Histograms of the infrared 
spectral index ($\alpha$, uncorrected for extinction) for the YSOs observed 
by \citet{oliveira2009:serpens}.  The shaded histogram shows the full sample 
of 78 objects whereas the dashed histogram shows the 20 objects identified as 
background AGB stars by Oliveira et al.  
{\it Right:} The black symbols and line plot the contamination fraction in 
each $\alpha$ bin, where this fraction is defined as the number of background 
AGB stars in this bin divided by the total number of objects in this bin, 
using the results from \citet{oliveira2009:serpens}.  The colored symbols plot 
the contamination fraction for each Class (red for Class 0+I, green for 
flat-spectrum, blue for Class II, and purple for Class III).  
The Class II and Flat-spectrum 
results are plotted at the mean value of $\alpha$ for each class, whereas 
the Class III and Class 0+I results are plotted at the mean values of the 
boundaries for each Class and the display ranges of the x axis.  
The dotted line in each panel marks the Class III boundary ($\alpha < -1.6$).}
\end{figure*}

Our final sample of YSOs is likely contaminated by background giant stars with 
infrared excesses.  \citet{oliveira2009:serpens} performed 
optical spectroscopy toward 78 YSOs in Serpens and identified 20 (26\%) as 
background Asymptotic Giant Branch (AGB) stars.  While the overall 
contamination rate is relatively 
low, Figure \ref{fig_contamination} shows that the contamination fraction 
strongly depends on the infrared spectral index $\alpha$\footnote{The quantity 
$\alpha$ is defined as the slope of the infrared SED in 
log($\lambda S_{\lambda}$) vs.~log($\lambda$) and is used to classify YSOs, 
as discussed in detail in \S \ref{sec_list} and \S \ref{sec_classification}.  
While $\alpha$ calculated from extinction corrected photometry is used 
in later sections, here we use the values calculated 
from the observed photometry for consistency with the previous studies to which 
we compare.}.  Starting from positive values, as $\alpha$ decreases, the 
contamination fraction remains close to zero 
until the Class II/III boundary is reached at $\alpha = -1.6$ (see \S 
\ref{sec_classification} for a full definition and discussion of YSO classes), 
and then steeply rises to $>$50\% for more negative values of $\alpha$.  On 
average, 62\% of the Class III sources examined by Oliveira et al.~are 
actually background AGB stars, whereas only 5\% of the Class II sources 
they examined were found to be contaminants.

Very high rates of AGB star contamination among Class III sources are also 
found by \citet{romero2012:transitiondisks}, also with optical spectra.  
They found Class III contamination rates of 100\% based on observations of 
30 objects in Lupus V and VI.  While the fact that they only observed objects 
that met initial selection criteria for transition disks may bias their 
results, they observed 45\% and 39\% of the total Lupus V and VI Class III 
populations, respectively; thus the true contamination fractions are at least 
$\sim$40\% in each cloud.

\begin{figure}
\epsscale{1.2}
\plotone{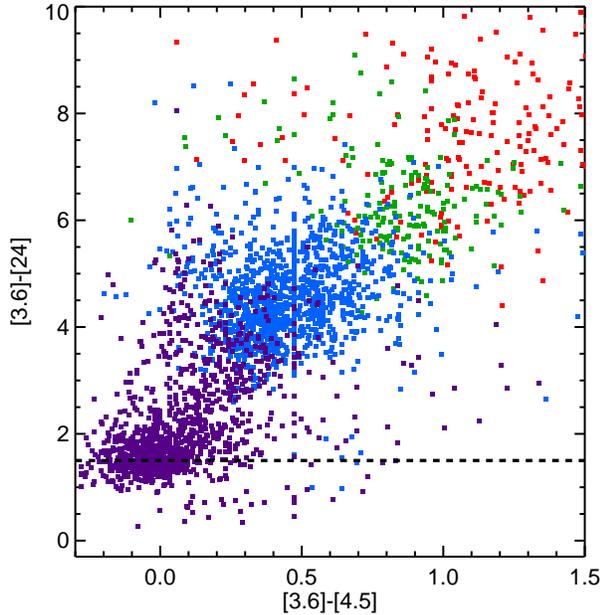}
\caption{\label{fig_cc_contamination}[3.6]--[24]~versus~[3.6]--[4.5] for all 
2966 objects in our final YSO sample, using the observed photometry.  The 
color of each symbol shows its spectral class determined using the extinction 
corrected infrared spectral index (see \S \ref{sec_list} and \S 
\ref{sec_classification} for details on classification), with red indicating 
Class 0+I, green indicating Flat-spectrum, blue indicating Class II, and purple 
indicating Class III.  The horizontal dashed line indicates the boundary of 
[3.6]--[24]~$=$~1.5 used to identify highly likely AGB stars contaminating our 
YSO sample.  The vertical line of sources at [3.6]--[4.5]~$\sim$~0.5 is 
a rounding artifact that is discussed in detail in Appendix 
\ref{sec_appendix_rounding}.}
\end{figure}

The true contamination fraction is likely a strong function of galactic 
latitude, and the specific clouds considered above (Serpens, Lupus V, and 
Lupus VI) all lie within $2-10$\degree\ of the galactic plane.  Indeed, 
\citet{romero2012:transitiondisks} found a lower contamination rate of 25\% 
for Ophiuchus ($b = 18$\degree), consistent with the Ophiuchus contamination 
rate of 20\% measured by \citet{cieza2010:transitiondisks} using similar 
methods.  To further investigate the contamination rate in our full sample 
of YSOs, Figure \ref{fig_cc_contamination} plots a \spitzer\ color-color 
diagram consisting of [3.6]--[24] versus [3.6]--[4.5], using the observed 
photometry.  Inspection of Figure \ref{fig_cc_contamination} shows 
that the Class III YSO population can be divided into a population of objects 
clustered at lower (bluer) 
values of both colors, and a second, more distributed 
population extending to higher (redder) colors.  Using optical spectra of 
candidate transition disks, \citet{cieza2010:transitiondisks}, 
\citet{cieza2012:transitiondisks}, and \citet{romero2012:transitiondisks} 
found that background AGB stars are predominately found at low values of 
[3.6]--[24], suggesting that the clustered population in Figure 
\ref{fig_cc_contamination} may be dominated by AGB stars whereas the more 
distributed population may be dominated by YSOs.  Indeed, the above studies 
found that all of their objects with [3.6]--[24]~$\leq$~1.5 were in fact AGB 
stars, with decreasing contamination rates at higher colors.  We find that 
298 of the objects in our final YSO sample have [3.6]--[24]~$\leq$~1.5, all 
but three of which are classified as Class III objects.  This represents 25\% 
of the Class III YSO sample and 10\% of the total YSO sample.  We consider 
these objects as likely AGB stars and mark them as such throughout the 
remainder of this paper.  However, we do not remove them from the sample 
since the studies cited above only targeted a subset of objects that were 
identified as transition disk candidates and thus there is no guarantee that 
their finding of 100\% contamination at the lowest values of [3.6]--[24] 
applies to our full sample.  Confirmation of the status of these objects 
requires either spectroscopic or proper motion studies.  Such work should be 
considered as a high priority for future investigations seeking to refine the 
YSO sample presented here.

\begin{figure}
\epsscale{1.2}
\plotone{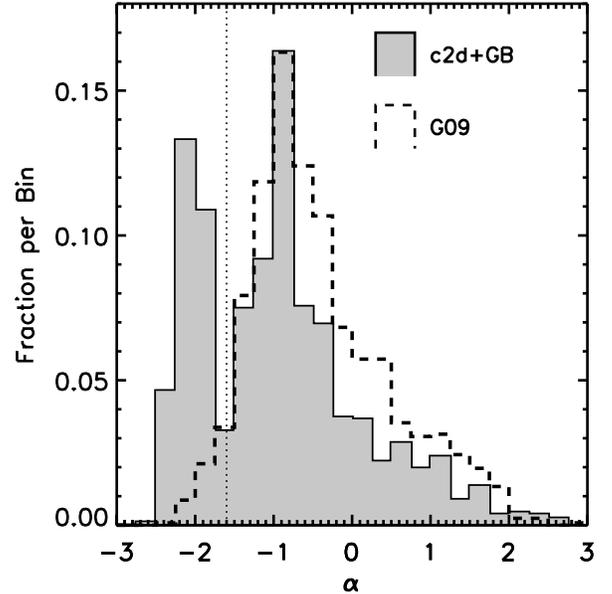}
\caption{\label{fig_gutermuth}Histogram showing the distribution of infrared 
spectral index for the YSOs identified in the c2d+GB clouds (this work, shaded 
histogram) and by \citet{gutermuth2009:ysos} in a survey of young stellar 
clusters (dashed histogram).  The dotted line marks the Class III boundary 
($\alpha < -1.6$).}
\end{figure}

The 25\% Class III contamination by AGB stars found above is almost certainly 
a lower limit to the full number, since some contaminating AGB stars are found 
to have [3.6]--[24]~$>$~1.5 \citep{cieza2010:transitiondisks,cieza2012:transitiondisks,romero2012:transitiondisks}.  However, these studies target samples 
that are too small and too biased to determine contamination rates as a 
function of [3.6]--[24] color.  
To obtain an upper limit to the possible contamination fraction, 
Figure \ref{fig_gutermuth} compares the distributions of $\alpha$ for our YSOs 
with those identified in young stellar clusters by \citet{gutermuth2009:ysos}, 
using a selection method that is also based on infrared colors and magnitudes 
but different in its details (see Gutermuth et al.~for details).  The c2d+GB 
sample contains a much higher fraction of Class III objects.  While a full 
comparison of different methods for identifying YSOs is beyond the scope of 
this work, under the assumptions that the Gutermuth et al.~method perfectly 
separates out background AGB stars and our regions contain the same fractions 
of sources in each Class, 87\% of the Class III objects in the c2d+GB sample 
would need to be removed to match the \citet{gutermuth2009:ysos} 
Class III fraction.  We note here that the Gutermuth et al.~survey targeted 
smaller regions centered on dense clusters whereas the c2d and GB surveys 
mapped much larger areas, including those more distant from cluster centers.  
As the Class II and III YSO populations are generally more distributed than 
earlier Classes, the intrinsic fraction of Class III sources is likely higher 
in the c2d and GB surveys than in \citet{gutermuth2009:ysos}.  For this reason, 
our above assumption that the surveys contain the same fractions of sources in 
each Class results in an upper limit to the Class III contamination fraction.

Synthesizing all of the above information, our sample of Class III YSOs 
suffers from between 25\% and 90\% contamination by AGB stars.  These 
values are lower and upper limits, respectively, for the reasons listed 
above, and the most likely value is somewhere in between.  Tighter constraints 
on the true contamination rate will require large optical spectroscopic 
campaigns and/or proper motion studies coupled with revisions to the selection 
criteria used to identify YSOs in infrared surveys.  These contamination rates 
must be kept in mind when considering the results presented here and in other 
studies based on the c2d+GB YSO list (such as those discussed in \S 
\ref{sec_intro}).

\subsection{Constructing SEDs and Correcting for Extinction}\label{sec_method_seds}

Following the method first summarized by \citet{evans2009:c2d} and updated by 
\citet{dunham2013:luminosities}, we compile as complete an SED as possible for 
each of the 2966 YSOs.  We refer to Dunham et al.~for a full description 
of this process, but give a summary here.  To construct the SEDs, we included: 
\begin{enumerate}
\item Selected ground-based optical and infrared photometry as compiled 
by the authors of the individual cloud papers listed in Table \ref{tab_clouds}.
\item \emph{Wide-field Infrared Survey Explorer} 
\citep[{\it WISE};][]{wright2010:wise} 12 and 22 \um\ 
photometry from the all-sky catalog\footnote{Available at http://irsa.ipac.caltech.edu/cgi-bin/Gator/nph-scan?mission=irsa\&submit=Select\&projshort=WISE}.
\item \emph{Spitzer} 160 \um\ photometry for sources that are detected but not 
located in saturated or confused regions.  Since the c2d pipeline does not 
extract 160 \um\ sources, we measured flux densities using aperture 
photometry (using the IDL procedure {\sc aper}) and standard aperture 
corrections as given by the MIPS Instrument 
Handbook\footnote{Available at http://irsa.ipac.caltech.edu/data/SPITZER/docs/ 
mips/mipsinstrumenthandbook/}.
\item Selected 350 \um\ photometry from a targeted survey of protostars with 
SHARC-II\footnote{The Submillimeter High Angular Resolution Camera II 
(SHARC-II) was a 350 \um\ bolometer array operated at the Caltech Submillimeter 
Observatory \citep{dowell2003:sharc}.} 
\citep[][A.~Suresh et al.~2015, submitted]{wu2007:sharc}.
\item Selected 450 and 850 \um\ photometry from the SCUBA\footnote{The 
Submillimeter Common-User Bolometer Array (SCUBA) was a 450 and 850 \um\ 
bolometer array operated at the James Clerk Maxwell Telescope.} Legacy Catalog 
\citep{difrancesco2008:scuba}.
\item Other submillimeter and millimeter photometry from the literature, where 
available from large surveys of nearby molecular clouds \citep{young2005:chamii,young2006:ophiuchus,enoch2006:bolocam,enoch2007:serpens,belloche2011:chami,belloche2011:chamiii,maury2011:aquila}.
\end{enumerate}

Regarding the last three items, we have access to complete submillimeter or 
millimeter surveys for only 6 out of the 18 clouds (Chamaeleon I, Chamaeleon 
II, Chamaeleon III, Ophiuchus, Perseus, and Serpens), plus a partial survey of 
Aquila and piecemeal coverage of other clouds from the SCUBA Legacy Catalog 
\citep{difrancesco2008:scuba} and our own 350 \um\ observations 
\citep[][A.~Suresh et al.~2015, submitted]{wu2007:sharc}.  We consider a 
\spitzer\ source to match a submillimeter or millimeter detection if it is 
located within one beam of the detection, using the appropriate beam size 
for each survey.  If multiple \spitzer\ sources match a single submillimeter 
or millimeter detection, we do not attempt to split the flux density and 
simply match it to the closest \spitzer\ source unless the SED clearly 
indicates a better match is obtained with a different \spitzer\ source.

Finally, we correct all photometry for foreground extinction following the 
procedure outlined by \citet{evans2009:c2d}.  We only wish to correct for 
the foregound extinction from both the clouds and the material between 
us and the clouds, and not the local extinction from dense cores surrounding 
protostars (this extincted emission is reprocessed to longer wavelengths and 
thus included in our compiled SEDs).  Thus, we adopt extinction values as 
follows:

\begin{enumerate}
\item Whenever possible, we adopt extinction values from the literature for 
Class II and III YSOs (classified via infrared spectral index; see \S 
\ref{sec_classification}), as determined from optical or near-infrared 
spectroscopic studies of each cloud that provide reliable spectral type 
(and thus extinction) measurements.  We refer to the data references 
listed in Table \ref{tab_clouds} for more details on these studies.
\item For all Class II and III YSOs with no published extinction values, we 
de-redden to the intrinsic near-infrared colors of an assumed spectral type of 
K7.  This particular spectral type is chosen because it is found to be 
representative of the YSOs in the c2d clouds \citep[][see also Evans et 
al.~2009 for details]{oliveira2009:serpens,oliveira2010:serpens}.
\item We calculate the mean extinction toward all Class II YSOs in each cloud, 
and then adopt that mean value for each of the Class 0+I and Flat-spectrum 
YSOs in that cloud.
\end{enumerate}

With extinctions assigned as above, we de-redden the SED of each YSO using the 
\citet{weingartner2001:dust} extinction law for $R_V = 5.5$.  Whereas 
\citet{evans2009:c2d} adopted a version of the \citet{weingartner2001:dust} 
$R_V = 5.5$ extinction law that only included dust absorption, 
here we update the procedure to correct for both dust absorption 
and scattering.  The choice of this particular extinction law is motivated 
by results showing that it is appropriate for the dense molecular clouds in 
which stars form \citep[e.g.,][]{chapman2009:dust}.  While we do caution that 
our approach only provides approximate extinction corrections, especially for 
the Class 0+I and Flat-spectrum sources where the corrections only account 
for the mean cloud extinction, it is nevertheless the best that can currently 
be done and is significantly more reliable than ignoring the effects of 
extinction altogether.

\section{Results}\label{sec_results}

\subsection{Full List of Young Stellar Objects}\label{sec_list}

For each of the 2966 YSOs, Table \ref{tab_ysos} lists a running index,
the cloud in which the YSO is found, the {\it Spitzer} source name (which
also gives the source coordinates), the visual extinction used for extinction
corrections, and the infrared spectral index ($\alpha$), the bolometric
temperature (\tbol), and the bolometric luminosity (\lbol).  These last three
quantities are calculated using both the observed and extinction corrected
photometry, and are denoted with primes in the latter case ($\alpha^{\prime}$,
\tbol$^{\prime}$, and \lbol$^{\prime}$).  The second-to-last column of Table 
\ref{tab_ysos} indicates whether or not each object has [3.6]--[24]~$\leq$~1.5 
and is thus likely a background AGB star, as described above in \S 
\ref{sec_method_contamination}.

The final column of Table \ref{tab_ysos} indicates whether or not each 
object is associated with a dense core as traced by submillimeter or 
millimeter continuum emission at $\lambda \geq 350$ \um.  As a lack of 
such an association does not necessarily imply that no core is present 
(see below), to avoid any confusion in the interpretation of this column we 
list either {\sc y} or no entry rather than {\sc y} or {\sc n}.  
\citet{dunham2013:luminosities} used such associations to define the sample 
of protostars in the c2d+GB clouds, with the rationale behind this definition 
being that detections in existing submillimeter and millimeter surveys of 
star-forming regions trace dense cores but not circumstellar disks due to 
the relatively low spatial resolutions and mass sensitivities of these surveys 
\citep[see][for details]{dunham2013:luminosities}.  By requiring a dense core 
but making no assumptions about the infrared colors of protostars, this 
definition ensures a reliable sample of protostars that includes those 
viewed edge-on down outflow cavities, but comes at the cost of incompleteness 
to those protostars that are embedded in low-mass cores or located in clouds 
with incomplete or no available submillimeter or millimeter continuum surveys 
\citep[see][for details]{dunham2014:ppvi}.  Associations with 
dense cores and the resulting sample of protostars must be revisited once the 
JCMT SCUBA-2 survey of the Gould Belt is complete, the first results of which 
are just starting to become available \citep[e.g.,][]{pattle2015:jcmtgb_oph}.

\begin{figure*}
\epsscale{0.85}
\plotone{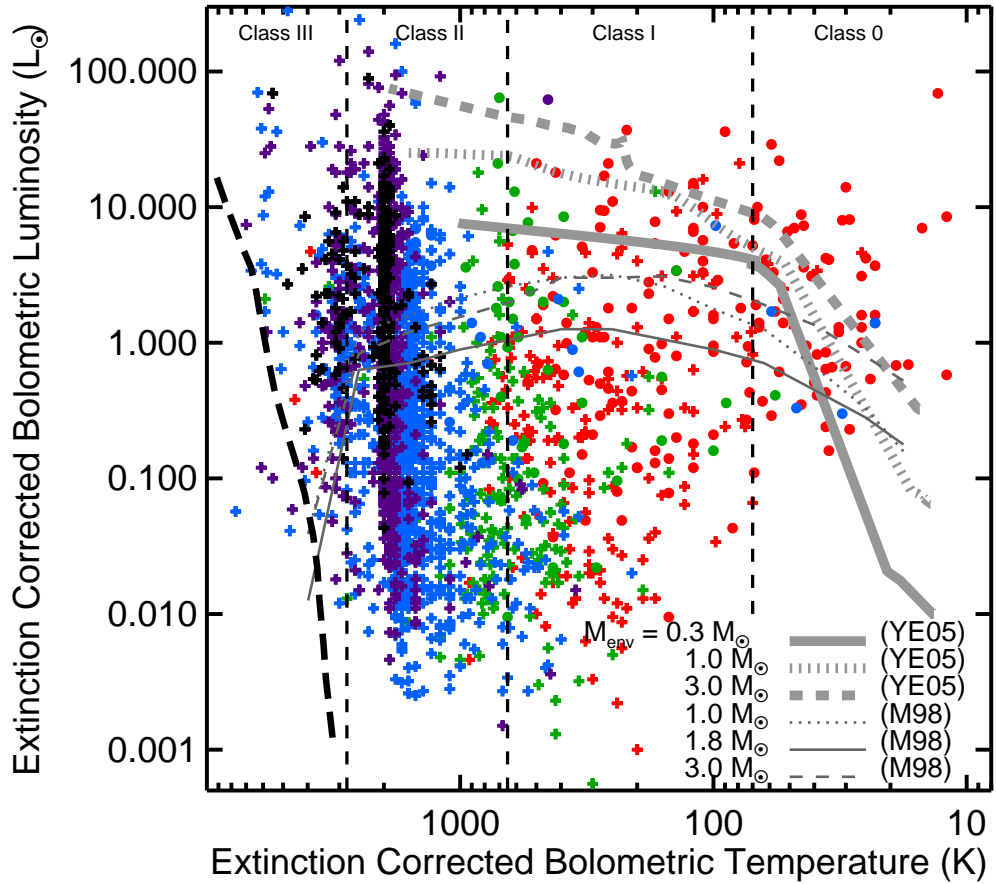}
\caption{\label{fig_blt}The extinction corrected bolometric luminosity 
(\lbolprime) plotted versus the extinction corrected bolometric temperature 
(\tbolprime) for all 2966 YSOs.  The color of each symbol shows its spectral 
class determined from \aprime\ (the extinction corrected infrared spectral 
index), with red indicating Class 0+I, green indicating Flat-spectrum, blue 
indicating Class II, and purple indicating Class III.  Black symbols indicate 
sources with [3.6]--[24]~$\leq$~1.5 that are considered likely to be background 
AGB stars (see \S \ref{sec_method_contamination}).  
Filled circles mark sources associated with submillimeter or millimeter 
continuum emission tracing dense cores (see text for details), while the plus 
signs mark sources with no such associations.  The three thick lines plot 
model tracks from \citet{young05:evolmodels} for singular isothermal spheres 
of different initial masses undergoing inside-out collapse with 100\% 
efficiency (all mass initially in the cores end up in the stars).  These 
models end when infall stops and the cores have fully accreted onto the stars 
and are thus only showing evolution in the protostellar stage.  The three 
thin lines plot model tracks from \citet{myers1998:evolmodels} for cores 
of different initial masses collapsing with accretion rates that decrease 
exponentially with time and efficiencies less than one \citep[i.e., only a fraction of the mass initially in the cores ends up in the stars, see][for details]{myers1998:evolmodels}.  As above, these models are only relevant for the 
protostellar stage of evolution.  The heavy dashed line on the left is the 
ZAMS from 0.1 to 2 \msun\ taken from \citet{dantona1994:zams}.  The \tbol\ 
class boundaries as defined by \citet{chen1995:tbol} are plotted as vertical 
dashed lines.}
\end{figure*}

The infrared spectral index ($\alpha$, $\alpha^{\prime}$) is defined as the 
slope of the infrared SED in log($\lambda S_{\lambda}$) vs.~log($\lambda$), 
where $\lambda$ is the wavelength and $S_{\lambda}$ is the flux density at that 
wavelength:

\begin{equation}\label{eq_alpha}
\alpha = \frac{d \, {\rm log}(\lambda S_{\lambda})}{d \, {\rm log}(\lambda)} \qquad .
\end{equation}
In practice, we calculate $\alpha$ with a linear least squares fit to all 
available 2MASS and {\it Spitzer} photometry between 2 and 24 \um.
The bolometric temperature of a source is defined to be the temperature of a 
blackbody with the same flux-weighted mean frequency \citep{myers1993:tbol}.  
It is in essence a protostellar equivalent of stellar effective temperature; 
it starts at very low values ($\sim 10$ K) for cold, starless cores and 
eventually increases to the effective temperature of the central (proto)star 
once the core and disk have fully dissipated.  \lbol, \lbol$^{\prime}$, \tbol, 
and \tbol$^{\prime}$ are calculated by integrating over the full SEDs:

\begin{equation}\label{eq_lbol}
\lbol = 4\pi d^2 \int_0^{\infty} S_{\nu}d\nu \qquad ,
\end{equation}
\begin{equation}\label{eq_tbol}
\tbol = 1.25 \times 10^{-11} \, \frac{\int_0^{\infty} \nu 
S_{\nu} d\nu}{\int_0^{\infty} S_{\nu} d\nu} \quad \rm{K} \qquad .
\end{equation}
In practice, we calculate these integrals using the trapezoid rule to integrate 
over the finitely sampled SEDs.  

Figure \ref{fig_blt} plots \lbolprime\ versus \tbolprime\ 
\citep[a ``BLT'' diagram;][]{myers1993:tbol} for all 2966 YSOs identified in 
the c2d and Gould Belt clouds.  The luminosities of Class 0 and Class I 
sources are distributed over greater than 4 orders of magnitude and extend 
to much lower luminosites than predicted by simple evolutionary models.  
A full discussion of this ``protostellar luminosity problem'' can be found 
in \citet{offner2011:luminosities}, \citet{dunham2013:luminosities}, and 
\citet{dunham2014:ppvi}, but to summarize, mass-dependent and/or time variable 
mass accretion is required to explain the observed luminosities of protostars.  
Class II and Class III objects exhibit both a very narrow range of 
bolometric temperatures and a bifurcation in \tbolprime\ into two groups.  
Both effects are artifacts introduced by our extinction corrections and 
are discussed in Appendix \ref{sec_appendix_tbol}.  To summarize, the first 
effect is because the bolometric temperature of a Class II or III source 
depends more on the spectral type of the object than the amount and 
characteristics of circumstellar material, and for most objects a constant 
spectral type of K7 is assumed (see above).  The second effect (bifurcation) 
is a consequence of optical data only being available for a subset of our 
YSOs.
 
Tables \ref{tab_fluxes_obs} and \ref{tab_fluxes_dered} list, for each 
YSO, the same running index as in Table \ref{tab_ysos}, followed by flux 
density and flux density uncertainty pairs for the 2MASS (1.25, 1.65, and 2.17 
\um) and {\it Spitzer} (3.6, 4.5, 5.8, 8.0, 24, and 70 \um) wavelengths.  
All flux densities and flux density uncertainties are listed in mJy and 
rounded to two significant digits.  
Table \ref{tab_fluxes_obs} lists the observed values while Table 
\ref{tab_fluxes_dered} lists the extinction corrected values.  The observed 
and extinction corrected flux densities that we have compiled at all other 
wavelengths in order to construct SEDS, as described above, are tabulated in 
Appendix \ref{sec_appendix_fluxes}.

\subsection{Classification of Young Stellar Objects}\label{sec_classification}

Over the past few decades a standard model for low-mass star formation has 
emerged.  This model, developed by \citet{adams1987:model} and summarized by 
\citet{shu1987:review}, resulted from the merger of an empirical classification 
system based on the infrared spectral indices of YSOs 
\citep{lada1984:classification} with theoretical studies of the stages of the 
collapse of a dense, rotating core \citep{shu1977:sis,terebey1984:model}.  
Recent reviews on classification of young stars and the associations between 
observed SED classes and physical evolutionary stages are given by 
\citet{evans2009:c2d} and \citet{dunham2014:ppvi}.

Including the revisions by \citet{greene1994:alpha}, these empirical Classes 
are as follows: 

\begin{itemize}
\item Class 0+I: \aprime\ $\geq 0.3$;
\item Flat-spectrum: $-0.3 \leq$ \aprime\ $< 0.3$;
\item Class II: $-1.6 \leq$ \aprime\ $< -0.3$;
\item Class III: \aprime\ $< -1.6$.
\end{itemize}
We denote the infrared spectral index as \aprime\ to 
emphasize that we only consider extinction corrected values in this study, 
and we refer to the first SED class as Class 0+I instead of the more commonly 
used Class I to emphasize that Class 0 and I objects are both included in 
this definition (see below).  

\begin{figure}
\epsscale{0.7}
\plotone{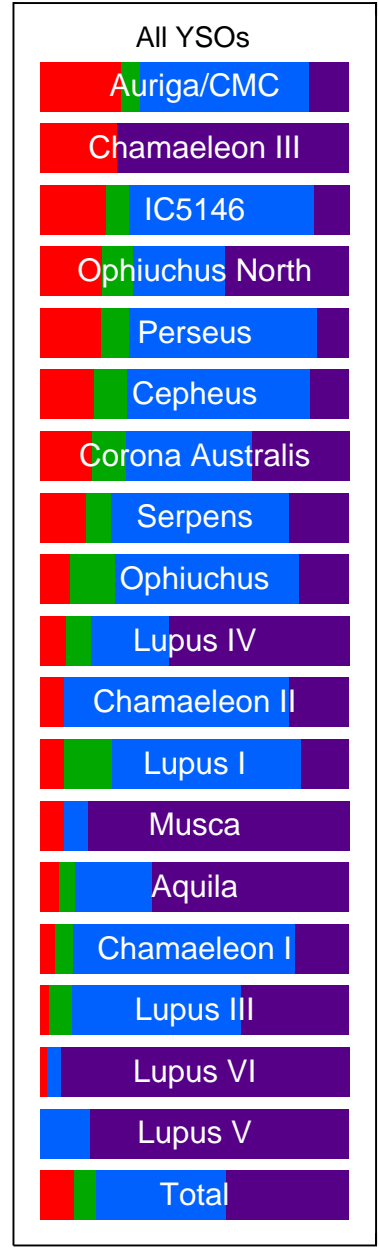}
\caption{\label{fig_classbars1}A four-color bar is plotted for each cloud, 
where each color corresponds to an evolutionary Class (similar to Figure 
\ref{fig_blt}, red corresponds to Class 0+I, green corresponds to 
Flat-spectrum, blue corresponds to Class II, and purple corresponds to 
Class III).  The fractional length of each color represents the fraction of 
YSOs in each cloud with the corresponding Class.  The clouds are sorted by 
Class 0+I fraction, with the largest fractions at the top and the smallest 
fractions at the bottom, and all 2966 YSOs are included.}
\end{figure}

\begin{figure}
\epsscale{0.7}
\plotone{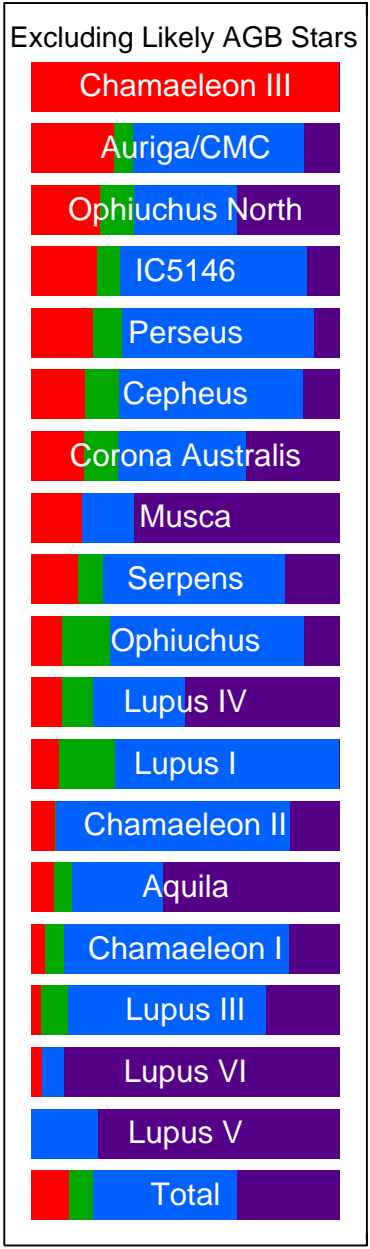}
\caption{\label{fig_classbars2}Same as Figure \ref{fig_classbars1}, except 
after removing the 298 objects identified as likely background AGB stars 
(see \S \ref{sec_method_contamination}).}
\end{figure}

Table \ref{tab_classification} lists, for each cloud, the numbers of YSOs 
in each Class using the extinction corrected infrared spectral indices, both 
with and without the likely background AGB stars removed.  Summed over all 
clouds, out of the 2966 total YSOs, 326 (11\%) are classified as Class 0+I, 
210 (7\%) are classified as Flat-spectrum, 1248 (42\%) are classified as 
Class II, and 1182 (40\%) are classified as Class III.  If we first remove 
the 298 sources identified as likely background AGB stars, out of the 
2668 remaining YSOs, 326 (12\%) are classified as Class 0+I, 210 (8\%) 
are classified as Flat-spectrum, 1245 (47\%) are classified as Class II, 
and 887 (33\%) are classified as Class III.  
We emphasize that there are large variations between clouds in the relative 
numbers in each Class.  To reinforce this point visually, Figures 
\ref{fig_classbars1} and \ref{fig_classbars2} plot, for each cloud, four-color 
bars where each color corresponds to a Class and the fractional length of each 
color represents the fraction of YSOs with the corresponding Class.  Figure 
\ref{fig_classbars1} displays the results including all 2966 YSOs whereas 
Figure \ref{fig_classbars2} displays the results after removing the 298 
likely background AGB stars.  Finally, we also 
emphasize that our method of YSO identification is only capable of identifying 
Class III sources with detectable infrared excesses and is thus incomplete to 
the full populations of pre-main sequence stars in these clouds.  

As noted above, we refer to the first Class as Class 0+I, but the original 
definition of this Class was simply Class I 
\citep{lada1984:classification,greene1994:alpha}.  Class 0 objects were later 
added as an earlier Class for protostars too deeply embedded to detect in the 
infrared but inferred through other means 
\citep[i.e., outflow presence;][]{andre1993:class0}.  They are defined as 
objects with a ratio of submillimeter to bolometric luminosity (\lsmmbol) 
greater than 0.5\% \citep{andre1993:class0}, where \lsmm\ is calculated for 
$\lambda \geq 350$ \um.  While they are defined observationally, the 
intended corresponding physical definition of Class 0 objects are protostars 
with at least 50\% of their total mass still residing in the surrounding, 
infalling core \citep{andre1993:class0}.  An alternative observational 
definition is given in terms of the bolometric temperature as objects with 
\tbol\ $< 70$ K \citep{myers1993:tbol,chen1995:tbol}.  While \lsmmbol\ is 
generally a better discriminator between Class 0 and Class I objects than 
\tbol, and is less sensitive to source geometry 
\citep[e.g.,][]{young05:evolmodels,dunham2010:evolmodels,dunham2014:ppvi}, 
it cannot be calculated without 350 \um\ photometry.  As we do not have 
350 \um\ photometry available for many of our YSOs, we only consider 
\tbol\ in this study.

\begin{figure}
\epsscale{1.2}
\plotone{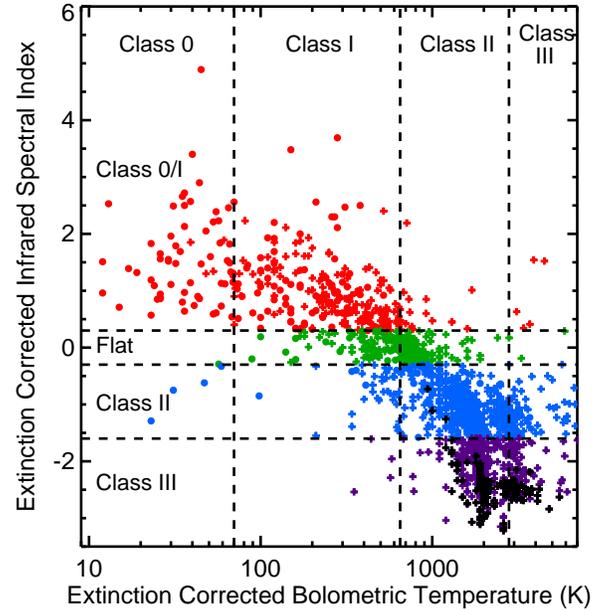}
\caption{\label{fig_alphatbol}Extinction corrected infrared spectral index 
(\aprime) plotted versus extinction corrected bolometric temperature 
(\tbolprime) for all 2966 YSOs, with the colors and symbols the same as in 
Figure \ref{fig_blt}.  The horizontal lines show the class boundaries in 
\aprime\ as revised by \citet{greene1994:alpha} and adopted in this work, 
and the vertical dashed lines show the \tbol\ class boundaries as defined 
by \citet{chen1995:tbol}.}
\end{figure}

With the sensitivity of {\it Spitzer}, Class 0 sources are easily detected 
in the infrared \citep[e.g.,][]{noriegacrespo2004:class0,young2004:l1014,jorgensen2006:perseus,dunham2006:iram04191,dunham2008:lowlum,tobin2010:protostars}.  
Figure \ref{fig_alphatbol} plots \aprime\ versus \tbolprime\ for all 2966 
YSOs.  In general, the two quantities are anti-correlated, although there are 
some notable exceptions.  There are a small number of SED Class 0+I sources 
(plotted in red in Figure \ref{fig_alphatbol}) with sufficiently high 
\tbolprime\ values to be classified as Class II or III by \tbolprime.  None of 
these objects are associated with dense cores and are likely simply more 
evolved YSOs viewed through large extinctions that are not fully 
extinction-corrected (recall that, for SED Class 0+I sources, only the average 
cloud extinction is accounted for).  There are also a few SED Class II YSOs 
(plotted in blue in Figure \ref{fig_alphatbol}) with sufficiently low 
\tbolprime\ values to be classified as Class 0 by \tbolprime.  These are likely 
true protostars viewed at nearly pole-on inclinations through outflow cavities 
\citep[e.g.,][]{whitney2003:geometry1,robitaille2006:models,crapsi2008:models}, 
although detailed follow-up study is required to rule out chance alignments 
between more evolved YSOs and dense cores.

While there is a general anti-correlation between \aprime\ and \tbolprime, 
this trend weakens for objects classified as Class 0+I by \aprime\ (objects 
with \aprime\ $\geq 0.3$).  With substantial overlap in \aprime\ for objects 
classified as Class 0 or Class I by \tbolprime, it is clear that these two 
Classes overlap in their infrared spectral indices.  These 
results confirm those presented previously by \citet{enoch2009:protostars} and 
\citet{evans2009:c2d} and indicate that mid-infrared data alone are 
insufficient to separate Class 0 and Class I objects; instead, 
full SEDs are required.  

Using our compiled SEDs, and restricting ourselves only to those YSOs 
associated with a dense core (both to restrict ourselves to confirmed 
protostars and to ensure available submillimeter or millimeter detections for 
reliable measurements of \tbolprime), our \tbolprime\ measurements result in 
65 Class 0 protostars and 129 Class I protostars.  Thus, we find that 
approximately one third ($65/[65+129]=0.335$) of protostars are classified as 
Class 0, in agreement with previous results based on the smaller c2d-only 
sample \citep{enoch2009:protostars,evans2009:c2d}.  
While future work must revisit 
this ratio once complete, sensitive submillimeter coverage is available for 
all clouds through the {\it Herschel} and JCMT Gould Belt surveys, 
preliminary \herschel\ results suggest only modest increases to the number of 
Class 0 objects \citep[generally 15\% -- 30\%,][]{harvey2013:auriga,stutz2013:pbrs,sadavoy2014:perseus}, leading to only small upward revisions in the ratio 
of Class 0 to Class I protostars.

\subsection{Infrared Color-Color Diagrams}\label{sec_cc}

\begin{figure*}
\epsscale{1.2}
\plotone{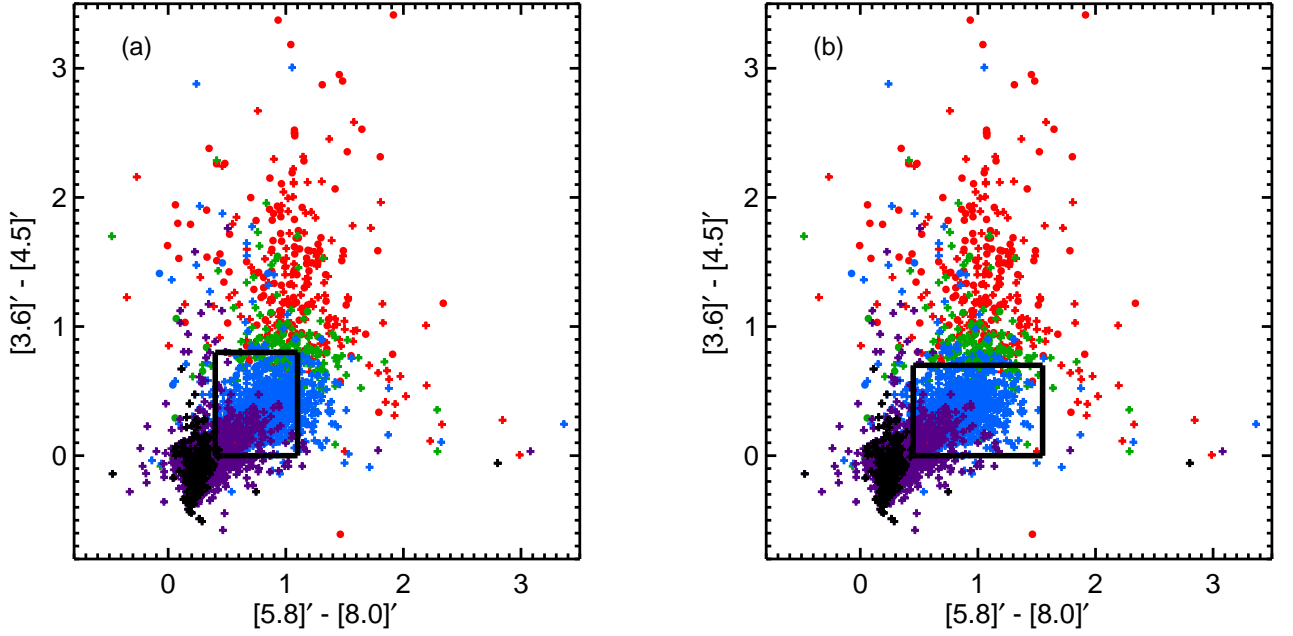}
\caption{\label{fig_cc1}Extinction corrected color-color diagrams for the four 
IRAC bands, with the colors and symbols the same as in Figure \ref{fig_blt}.  
The box in (a) indicates the region identified with Class II sources by 
\citet{allen2004:irac}, whereas in (b) it indicates our proposed modifications 
to this region (see text for details).}
\end{figure*}

Figure \ref{fig_cc1} presents an extinction corrected 
infrared color-color diagram of all YSOs 
identified in this work constructed from the four IRAC wavelengths: 
[3.6]$^{\prime}$~$-$~[4.5]$^{\prime}$ plotted versus 
[5.8]$^{\prime}$~$-$~[8.0]$^{\prime}$.  The different Classes of 
YSOs (classified via \aprime) are generally well-separated in this diagram, 
as first noted by \citet{allen2004:irac}.  In cases where photometry is only 
available at $3.6-8.0$ \um\ and thus the infrared spectral index cannot be 
reliably calculated, this diagram can be used to approximately classify YSOs.  
Indeed, \citet{allen2004:irac} used the following two conditions to define a 
box where Class II sources are located:
\begin{enumerate}
\item 0.4 $\leq$ [5.8]~$-$~[8.0] $\leq$ 1.1, and 
\item 0.0 $\leq$ [3.6]~$-$~[4.5] $\leq$ 0.8.
\end{enumerate}
The original definition of this ``Class II box'' by \citet{allen2004:irac} 
used observed photometry.  Inspection of Figure \ref{fig_cc1} reveals that 
minor revisions are required when using extinction corrected photometry.  
While we acknowledge that polygons with greater than four sides could 
possibly give better results, for simplicity we only consider rectangular 
regions.  To define a 
new Class II box, we allow all four boundaries to vary and select those which 
simultaneously maximize the probability that a Class II YSO is located within 
the box (defined as the number of Class II YSOs located within the box divided 
by the total number of Class II YSOs) and the probability that a YSO located 
within the box is classified as Class II (defined as the number of Class II 
YSOs located within the box divided by the total number of YSOs located within 
the box).  In practice, we implement these definitions 
by selecting the boundaries which 
maximize the sum of these two probabilities.  Our proposed new Class II box, 
plotted in panel (b) of Figure \ref{fig_cc1}, has the following boundaries:
\begin{enumerate}
\item 0.45 $\leq$ [5.8]$^{\prime}$~$-$~[8.0]$^{\prime}$ $\leq$ 1.55, and 
\item 0.00 $\leq$ [3.6]$^{\prime}$~$-$~[4.5]$^{\prime}$ $\leq$ 0.70.
\end{enumerate}
We note that the shape of this revised box is qualitatively similar to that 
proposed by \citet{gutermuth:phd}, who followed a similar procedure.  
With these revised boundaries, the fraction of Class II YSOs located within 
the box increases from 84.5\% to 89.9\%, and the fraction of YSOs in the box 
that are Class II increases from 78.8\% to 82.4\%.

\begin{figure*}
\epsscale{1.2}
\plotone{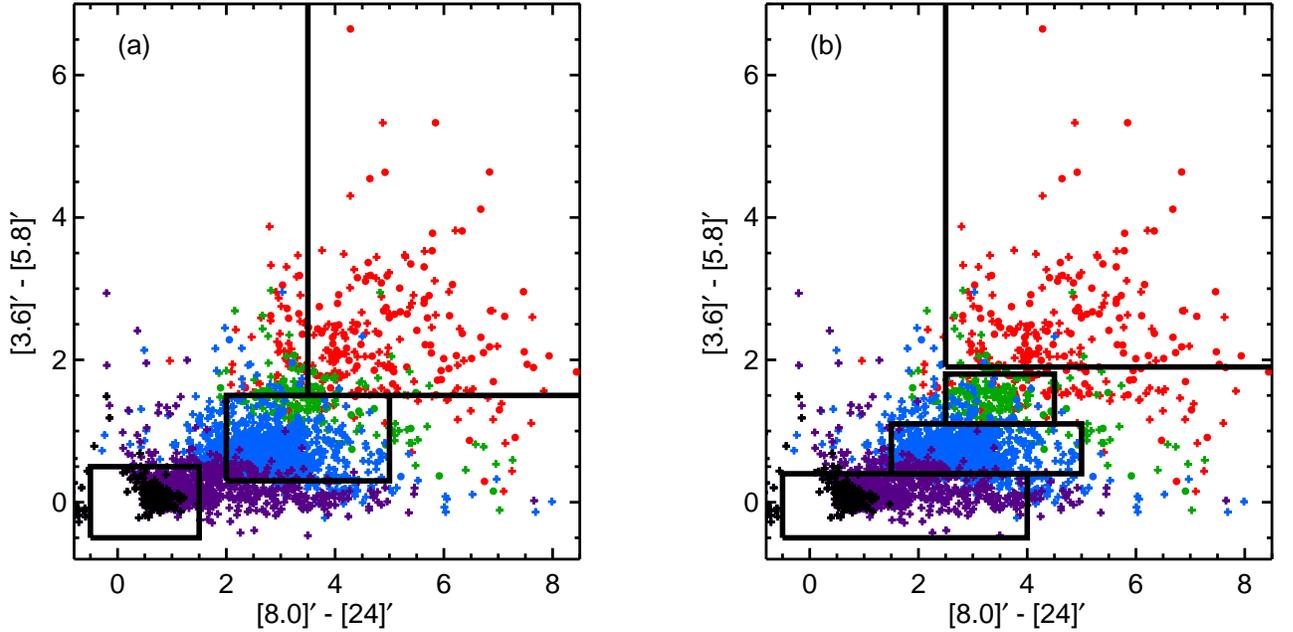}
\caption{\label{fig_cc2}Extinction corrected 
color-color diagram using three of the four IRAC bands 
and the first MIPS band, with the colors and symbols the same as in Figure 
\ref{fig_blt}.  The boxes in (a) indicate the regions identified with 
Class III/stellar, Class II, and Class 0+I sources by 
\citet{muzerolle2004:spitzer}, going from lower left to upper right.  
The boxes in (b) indicate our proposed modifications to these regions, 
now with four regions (Class III/stellar, Class II, Flat-spectrum, 
and Class 0+I, again from lower left to upper right).}
\end{figure*}

Figure \ref{fig_cc2} presents a second extinction corrected infrared 
color-color diagram of all YSOs which 
uses both the IRAC and MIPS observations by plotting 
[3.6]$^{\prime}$~$-$~[5.8]$^{\prime}$ versus 
[8.0]$^{\prime}$~$-$~[24]$^{\prime}$.  Again the different Classes of YSOs 
(classified via \aprime) 
are generally well-separated, as first noted by \citet{muzerolle2004:spitzer}.  
Indeed, Muzerolle et al.~defined three regions in this color-color diagram 
to separate Class III/stellar, Class II, and Class 0+I YSOs; these regions are 
marked in panel (a) of Figure \ref{fig_cc2} and are defined in Table 
\ref{tab_cc2}.  As above, Muzerolle et al.~used observed photometry, and we 
find that minor revisions are required when using extinction corrected 
photometry.  Furthermore, the original boundaries 
defined by Muzerolle et al.~did not include a region for Flat-spectrum objects. 
Thus, we follow the same procedure as above, again only considering 
rectangular regions, and maximize the sum of the eight 
probabilities (two for each Class), subject to the following constraints: 
(1) the minimum x and y Class III boundaries be held fixed at the 
Muzerolle et al.~values, (2) there are no maximum x and y Class 0+I boundaries, 
and (3) the regions can not overlap.  Our revised 
boundaries are plotted in panel (b) of Figure \ref{fig_cc2} and listed in 
Table \ref{tab_cc2}.  Table \ref{tab_cc2} also lists each of the two 
probabilities considered for both the original and revised boundaries.

Ultimately, no region in either color-color diagram will classify YSOs with 
a perfect one-to-one correspondence to the infrared spectral index, but in 
cases where insufficient photometry at 2 and/or 24 \um\ is available, 
approximate YSO classification is possible via infrared color-color diagrams.

\section{Timescales for Young Stellar Objects}\label{sec_lifetimes}

\subsection{Class 0+I and Flat-spectrum YSOs}

The standard method for determining the time spent in each YSO Class is to 
calculate the ratio of the number of sources in that Class to the number 
in a reference Class, and then multiply by the duration of the reference 
Class \citep[e.g.,][]{wilking1989:oph,evans2009:c2d,dunham2014:ppvi}.  
Implicit in this method are the assumptions that star formation is 
continuous over at least the duration of the reference class and that time 
is the only variable.  Regarding the first assumption, we strive to average 
out cloud-to-cloud variations in the recent star formation history by 
averaging over many clouds.  Regarding the second, our 
results are best thought of as an average over all other relevant variables 
\citep[e.g., environment, stellar mass; see][]{evans2009:c2d}.  
In this section, we investigate the effects of different assumptions for 
the reference class and its duration on the timescales of Class 0+I and 
Flat-spectrum YSOs.

We first follow the method of \citet{evans2009:c2d} and 
\citet{dunham2014:ppvi}, who adopt Class II as the reference Class and assume 
a Class II duration of 2 Myr.  This duration is taken from analyses of the 
fraction of cluster members with disks versus cluster age for various 
young clusters, which generally show a disk disperal time of a few Myr 
\citep[][]{wyatt2008:disks,ribas2014:ages,ribas2015:ages}.  As discussed 
extensively by \citet{evans2009:c2d}, this time is best considered as a disk 
half-life rather than an absolute duration.  The first row of Table 
\ref{tab_timescales} presents durations for the Class 0+I and Flat-spectrum 
Classes calculated following this method, using the extinction corrected 
numbers: 0.52 and 0.34 Myr for Class 0+I and Flat-spectrum objects, 
respectively.

Although the ages of young stars are highly uncertain, recent studies suggest 
a revised half-life for disk survival of $2.5-4$ Myr on both observational 
\citep[e.g.,][]{kraus2012:ages,bell2013:lifetimes,soderblom2014:ppvi} and 
theoretical \citep[e.g.,][]{alexander2009:ages} grounds.  Very recent results 
by \citet{ribas2014:ages} and \citet{ribas2015:ages}, using spectroscopically 
confirmed members of various young clusters,  show that this duration 
depends on the wavelength where the infrared excess indicative of the presence 
of a disk first appears, ranging from 1.9 Myr for $\lambda < 5$ \um\ to 
3.0 Myr for $\lambda \leq 24$ \um\ (note that these times have been converted 
from the $1/e$ times presented by Ribas et al.~to half-life times).  As our 
\spitzer\ data are sensitive to infrared excesses out to 24 \um, and our 
\spitzer\ c2d+GB data have comparable mid-infrared sensitivities to the 
Ribas et al.~{\it Spitzer} and {\it WISE} data used to determine disk 
lifetimes, their derived lifetimes suggest that 3 Myr is a more appropriate 
choice for the duration of the reference Class in this work.  The second row 
of Table \ref{tab_timescales} lists the revised durations: 0.78 and 0.50 Myr 
for Class 0+I and Flat-spectrum objects, respectively.

While it has become somewhat standard to adopt Class II as the reference 
Class, this 
choice may result in a logical inconsistency.  The duration of either 2 
or 3 Myr based on disk fraction vs.~cluster age is derived by determining 
the ratio of cluster members with infrared excesses to all cluster members 
(in other words, the ratio of disk sources to disk $+$ no disk sources).  
By adopting Class II as the reference Class, we omit all of the Class III 
YSOs that still have disks, even though these objects are included when other 
studies (such as those by Ribas et al.~) derive disk lifetimes.  Since 
infrared colors indicative of dust are the fundamental ingredient in our 
identifcation of YSOs (see \S \ref{sec_method_yso}), by definition the Class 
III YSOs identified in this work have disks (if they are all in fact YSOs; 
the issue of Class III contamination will be addressed in the following 
paragraph).  Since the recent work by \citet{ribas2014:ages} and 
\citet{ribas2015:ages} is based on mid-infrared data from {\it Spitzer} and 
{\it WISE} with comparable sensitivity to mid-infrared excesses as our 
c2d+GB data, and since we use disk lifetimes derived by Ribas et al.~to set 
the duration of the reference Class, and further given that all Class II and 
Class III sources identified here have disks, we thus argue that it is more 
appropriate to adopt the sum of Class II and Class III as the reference Class 
than Class II alone.  Thus, the third row of Table \ref{tab_timescales} 
lists the revised durations assuming a duration of 3 Myr for the 
sum of Class II and III sources: 0.40 and 0.26 Myr for Class 0+I and 
Flat-spectrum objects, respectively.

The one major complication with the above analysis is the fact that our 
Class III sample suffers from 25\% -- 90\% contamination, as discussed in 
\S \ref{sec_method_contamination}.  On the other hand, contamination is not 
a problem in the \citet{ribas2014:ages,ribas2015:ages} samples since they 
only consider spectroscopically confirmed cluster members.  As a result, the 
above calculations of Class durations likely overcount the number of objects 
in the reference Class and thus underestimate the Class 0+I and Flat-spectrum 
durations.  To account for this range of 
contamination rates, the last two columns of Table \ref{tab_timescales} list 
the calculated durations of the Class 0+I and Flat-spectrum Classes with the 
Class II + Class III reference Class corrected for contamination and a 
reference Class duration of 3 Myr.  We consider these values to be our most 
accurate and logically consistent lifetimes: $0.46-0.72$ Myr for Class 0+I 
sources, and $0.30-0.46$ Myr for Flat-spectrum sources.

\subsection{Class 0 YSOs}\label{sec_timescales_class0}

In \S \ref{sec_classification}, we reported that 33.5\% of the objects 
associated with dense cores (and thus considered protostars according to 
the definition adopted in this paper) are classified 
as Class 0 by \tbolprime, with the remainder classified as Class I.  
Under the same assumptions as above (continuous star formation and time is the 
only variable) and our definition of a protostar as a YSO associated with 
a dense core, our results imply that 33.5\% of the total 
protostellar duration is spent in the Class 0 phase, with the remaining 66.5\% 
of the time spent in the Class I phase.  If we assume that the 
lifetime of the SED Class 0+I objects is a good proxy for the lifetime of the 
protostellar stage of evolution (see \S \ref{sec_timescales_protostars} below 
for justification of this assumption), the range of SED Class 0+I durations 
listed in Table \ref{tab_timescales} ($0.40-0.78$ Myr) imply resulting 
durations of $0.13-0.26$ Myr for Class 0 protostars and $0.27-0.52$ Myr for 
Class I protostars (see the last two columns of Table \ref{tab_timescales}).  
If we only consider the last two rows of Table \ref{tab_timescales} (including 
Class III YSOs in the reference Class, correcting for estimated contamination 
rates in the Class III sample, and adopting a reference Class duration of 3 
Myr), the resulting durations are $0.15-0.24$ Myr for Class 0 and $0.31-0.48$ 
Myr for Class I.  

As noted in \S \ref{sec_classification}, we generally find the same ratios of 
SED Class 0 to SED Class 0+I YSOs as in previous studies 
\citep{evans2009:c2d,enoch2009:protostars,dunham2014:ppvi}.  While 
those studies derived a Class 0 duration of $0.15-0.17$ Myr, our expanded 
range of possible durations is due to our expanded analysis of the contents 
and duration of the reference Class.

\subsection{Lifetimes of Protostars}\label{sec_timescales_protostars}

As noted in \S \ref{sec_classification}, the standard picture of low-mass 
star formation resulted from the merger of an empirical classification 
system based on the infrared spectral indices of YSOs 
\citep{lada1984:classification} with theoretical studies of the stages of the 
collapse of a dense, rotating core \citep{shu1977:sis,terebey1984:model}.  
In this picture, SED Class 0+I YSOs are protostars still embedded in and 
accreting from dense cores, SED Class II YSOs are T Tauri stars with 
optically thick, accreting disks, and SED Class III YSOs are pre-main 
sequence stars with little or no disks left (all {\it Spitzer}-identified 
Class III YSOs, by definition, still have infrared excesses, thus the 
population of Class III YSOs without disks is not identified in the work 
presented here).  Flat-spectrum YSOs have no clear link to a distinct 
physical stage, and in reality the correlation between observational SED 
Class and physical Stage is weakened by the effects of geometry, extinction, 
and accretion history \citep[e.g.,][]{whitney2003:geometry1,young05:evolmodels,robitaille2006:models,crapsi2008:models,dunham2010:evolmodels}.

While our study provides the statistics necessary to measure durations of 
the observed SED Classes, it does not provide the necessary information 
to measure durations of physical Stages.  In a separate, complementary study, 
\citet{heiderman2015:misfits} surveyed nearly all of the SED Class 0+I 
and Flat-spectrum YSOs in the c2d and GB surveys for emission from the dense 
gas tracer \hcopjthree.  
Using criteria based on the detection and strength of this line, 
they sorted objects into Stage 0+I (protostar) and Stage II (disk) sources.  
They found that the Flat-spectrum YSOs are approximately evenly divided between 
protostar and disk sources, with no evidence that they occupy a physically 
distinct evolutionary stage.  They also found that the number of Flat-spectrum 
YSOs classified as protostars is balanced by the number of SED Class 0+I 
YSOs classified as disks, coincidentally resulting in a nearly identical 
number of protostars (defined in their work based on the presence of 
dense gas traced by \hcop\ line emission) and SED Class 0+I YSOs (326 
Class 0+I YSOs identified here, 335 protostars identified by Heiderman \& 
Evans).  Thus, based on their results, the range of possible 
protostellar stage durations is the same as the range of possible SED 
Class 0+I durations listed in Table \ref{tab_timescales}.

\section{Summary}\label{sec_summary}

In this paper we have presented the full catalog of young stellar objects 
identified in the molecular clouds surveyed by the {\it Spitzer Space 
Telescope} ``cores to disks'' (c2d) and ``Gould Belt'' (GB) Legacy surveys.  
In the case of the c2d clouds, our catalog updates a previous one 
published by \citet{evans2009:c2d}, and in the case of the GB clouds, our 
catalog represents the first catalog of {\it Spitzer}-identified YSOs in the 
full ensemble of these clouds.  We summarize our main results as follows:

\begin{itemize}
\item We have used standard techniques developed by the c2d project to identify 
3239 candidate YSOs in the 18 c2d and GB clouds, 2966 of which survive visual 
inspection and form our final catalog of young stellar objects in the Gould 
Belt.
\item We compile SEDs for all 2966 YSOs and correct these SEDs for 
foreground extinction.
\item We calculate and tabulate the infrared spectral index, bolometric 
luminosity, and bolometric temperature for each object, both with and without 
extinction corrections.  We also tabulate information on whether or not each 
YSO is considered highly likely to be a background AGB star, and whether or 
not each YSO is associated with a dense core as traced by submillimeter or 
millimeter continuum emission.
\item After correcting for extinction, we find that 326 out of 2966 (11\%) are 
classified as Class 0+I, 210 out of 2966 (7\%) are classified as Flat-spectrum, 
1248 out of 2966 (42\%) are classified as Class II, and 1182 out of 2966 (40\%) 
are classified as Class III.  The Class III sample suffers from an overall 
contamination rate by background AGB stars between 25\% and 90\%.
\item Adopting standard assumptions, we derive durations of $0.40-0.78$ Myr 
for Class 0+I YSOs and $0.26-0.50$ Myr for Flat-spectrum YSOs, with most 
likely durations of $0.46-0.72$ Myr (Class 0+I) and $0.30-0.46$ Myr 
(Flat-spectrum).  The ranges encompass uncertainties in the adopted assumptions.
\item Including information from submillimeter and millimeter wavelengths, 
one-third of the Class 0+I sample is classified as Class 0.  With this result, 
we calculate durations of $0.13-0.26$ Myr (Class 0) and $0.27-0.52$ Myr 
(Class I), with most likely durations of $0.15-0.24$ Myr (Class 0) and 
$0.31-0.48$ Myr (Class I).  As above, the ranges encompass uncertainties in 
the adopted assumptions.
\item We revisit infrared color-color diagrams used in the literature to 
classify YSOs and use our large, extinction corrected sample to propose 
revisions to classification boundaries in these diagrams.
\item The bolometric temperature of a YSO surrounded by a circumstellar disk 
but no longer embedded in a dense core is primarily determined by the spectral 
type of the YSO and the availability of optical photometry, and not by the 
properties of the circumstellar material.  As such, the bolometric temperature 
should not be used to distinguish between Class II and Class III YSOs.
\end{itemize}

The YSO catalog presented here represents the most complete catalog of 
star formation within 500 pc of the Sun yet assembled.  It includes nearly 
every major site of star formation in the solar neighborhood except for the 
Taurus and Orion molecular clouds, each of which was the subject of its own, 
dedicated {\it Spitzer} Legacy survey.  While this work represents a major 
step forward in assembling a complete and reliable catalog of Young Stellar 
Objects in the Gould Belt, future work to further refine the YSO 
catalog presented here is needed on the following three fronts:  (1) combining 
the results of the c2d+GB, Taurus, and Orion surveys using uniform data 
reduction and analysis techniques, (2) identifying and removing background 
AGB stars currently contaminating the sample, and (3) revisiting the 
associations between \spitzer\ sources and dense cores traced by submillimeter 
and millimeter detections once the full results from the \herschel\ and SCUBA-2 
Gould Belt surveys are available.

\acknowledgments
We thank the referee for thoughtful comments that have improved 
the quality of this paper, and we acknowledge helpful discussions with 
H.~Arce and L.~Kristensen.  
This work is based primarily on observations obtained with the \emph{Spitzer 
Space Telescope}, operated by the Jet Propulsion Laboratory, California 
Institute of Technology.  It also uses data products from the Two Micron All 
Sky Survey, which is a joint project of the University of Massachusetts and 
the Infrared Processing and Analysis Center/California Institute of 
Technology, funded by the National Aeronautics and Space Administration and 
the National Science Foundation.  These data were provided by the NASA/IPAC 
Infrared Science Archive, which is operated by the Jet Propulsion Laboratory, 
California Institute of Technology, under contract with NASA.  This research 
has made use of NASA's Astrophysics Data System (ADS) Abstract Service, 
the IDL Astronomy Library hosted by the NASA Goddard Space Flight Center, and 
the SIMBAD database operated at CDS, Strasbourg, France.  
MMD acknowledges support from NASA ADAP grant NNX13AE54G and 
from the Submillimeter Array (SMA) through an 
SMA postdoctoral fellowship and from H.~Arce at Yale University, where a 
portion of this work was performed.   NJE is supported by NSF Grant 
AST-1109116 to the University of Texas at Austin.  LAC was supported by 
ALMA-CONICYT grant 31120009, CONICYT-FONDECYT grant 1140109, and the 
Millennium Science Initiative (Chilean Ministry of Economy) grant RC130007.  
ALH is supported by an NSF AAPF under award AST$-$1302978.  
KEY acknowledges support from the Louisiana Space Consortium Research 
Enhancement Award through NASA EPSCoR grant number NNX10AI40H.

\bibliographystyle{apj.bst}
\bibliography{dunham_citations}

\appendix

\section{A.~~{\it Spitzer} Astronomical Observation Requests (AORs) for Unpublished Clouds}

Three clouds listed in Table \ref{tab_clouds} do not have published data 
references (Chamaeleon I, Chamaeleon III, and Musca).  Additionally, only a 
small piece of Aquila was published in the reference listed for that 
cloud \citep{gutermuth2008:serpsouth}.  Consequently, for these four clouds 
there are no published lists of the \spitzer\ Astronomical Observation 
Requests (AORs) that targeted them.  We thus list their AORs in Table 
\ref{tab_appendix_aors}.

\section{B.~~Rounding Artifacts in Color-Color Diagrams}\label{sec_appendix_rounding}

Inspection of Figure \ref{fig_cc_contamination} reveals a vertical line of 
sources at [3.6]--[4.5] $\sim$ 0.5.  If an object has identical flux 
densities at two {\it Spitzer} wavelengths, the magnitude difference between 
these two wavelengths is then simply equal to:

\begin{equation}
m_1 - m_2 = 2.5 \, {\rm log} \, \frac{ffzm_2}{ffzm_1} \qquad ,
\end{equation}
where $ffzm_1$ and $ffzm_2$ are the fluxes for zero magnitude at each 
wavelength. For 3.6 and 4.5 \um, this gives [3.6]--[4.5]~=~0.473.

\begin{figure}
\epsscale{0.6}
\plotone{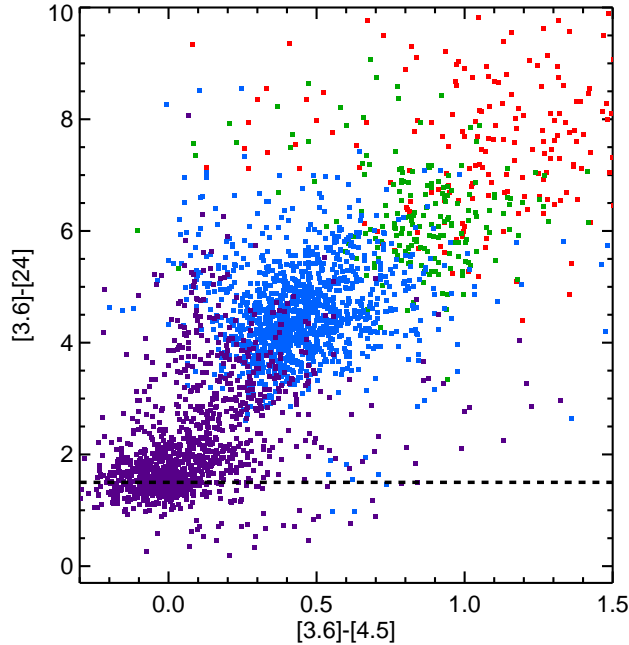}
\caption{\label{fig_appendix_cc}Same as Figure \ref{fig_cc_contamination} 
([3.6]--[24]~versus~[3.6]--[4.5] for all 2966 objects in our final YSO sample, 
using the observed photometry), except plotted using the original data before 
rounding.  The vertical line of sources at [3.6]--[4.5] $\sim$ 0.5 has now 
disappeared as YSOs with similar flux densities have not now all been rounded 
to identical flux densities.}
\end{figure}

There are 147 objects plotted in Figure \ref{fig_cc_contamination} with 
identical flux densities at 3.6 and 4.5 \um, resulting in 147 objects plotted 
at the same value on the x-axis and thus giving rise to the vertical line 
of sources.  We note that this is an artifact introducing by rounding 
all flux densities to two significant digits.  To demonstrate this point, 
Figure \ref{fig_appendix_cc} plots the same color-color diagram as Figure 
\ref{fig_cc_contamination}, except now using the original data before 
rounding.  As evident from inspection of this figure, the vertical line of 
sources at [3.6]--[4.5] $\sim$ 0.5 has now disappeared.

The same artifact is also present in Figure \ref{fig_cc_contamination} in the 
[3.6]--[24] color.  For 3.6 and 24 \um, identical flux densities result in a 
magnitude difference of [3.6]--[24]~=~3.959.  
There are only 23 objects plotted in Figure 
\ref{fig_cc_contamination} with identical flux densities at 3.6 and 24 \um, 
so while the vertical line of sources at [3.6]--[24] $\sim$ 4 is present, 
it is not as apparent.  Similar artifacts also affect the colors 
plotted in Figures \ref{fig_cc1} and \ref{fig_cc2} using the extinction 
corrected photometry.  For the four colors plotted in these two figures, 
between 38 and 95 objects with identical flux densities are plotted.  While 
corresponding vertical and horizontal lines of sources are present, none 
are apparent because they all occur at colors where the source densities are 
too high for these small subsets of sources to be visible.  These rounding 
artifacts have no effect on the results presented in this paper.

\section{C.~~Bolometric Temperatures of Class II and III Sources}\label{sec_appendix_tbol}

As noted in \S \ref{sec_list}, the bolometric temperatures of our Class II and 
III YSOs are generally tightly clustered into two narrow \tbolprime\ ranges 
(see Figures \ref{fig_blt} and \ref{fig_alphatbol}). To understand 
the origins of these effects, we construct simple continuum radiative 
transfer models of circumstellar disks using the three-dimensional Monte Carlo 
radiative transfer package RADMC-3D\footnote{Available at: http://www.ita.uni-heidelberg.de/$\sim$dullemond/software/radmc-3d/}.  
We assume an axisymmetric physical 
structure and use RADMC-3D to first calculate the two-dimensional dust 
temperature profile through the disks and then generate full model SEDs from 
optical to millimeter wavelengths.

We adopt a simple, parameterized disk density structure following a power-law 
in the radial coordinate and a Gaussian in the vertical coordinate.  The 
exact density profile is given as follows:
\begin{equation}\label{eq_disk_density_profile}
\rho_{disk}(s,z) = \rho_0 \left(\frac{s}{s_o} \right)^{-\alpha} 
exp\left[ -\frac{1}{2} \left(\frac{z}{H_s}\right)^2 \right] \qquad ,
\end{equation}
where $z$ is the distance above the midplane ($z=rcos\theta$, with $r$ and 
$\theta$ the usual radial and zenith angle spherical coordinates), $s$ is the 
distance in the midplane from the origin ($s= \sqrt{r^2 - z^2}$), and 
$\rho_0$ is the density at the reference midplane distance $s_0$, 
normalized such that the total disk mass is correct (see below).    
The quantity $H_s$ is the disk scale height and is given by 
$H_s = H_o \left(\frac{s}{s_o}\right)^\beta$, where $H_o$ is the scale height 
at $s_o$.  The parameter $\beta$, which describes how the scale height 
changes with $s$ and sets the flaring of the disk, is treated as a free 
parameter (see below).  We set $H_o = 10$ AU at $s_o = 100$ AU.  
The disk surface density profile, calculated by 
integrating Equation \ref{eq_disk_density_profile} over the vertical 
coordinate $z$, has a radial power-law index of $\Sigma(s) \propto s^{-p}$, 
where $p = \alpha - \beta$.  For these models, we hold $\alpha$ fixed at 2.25 
and the disk outer radius fixed at 100 AU.

To consider a diverse set of disks, we treat the disk inner radius, 
flaring parameter $\beta$, and total disk mass as free parameters.  Allowed 
values for each parameter are as follows:  0.1, 1, 10, 20, and 50 AU for the 
disk inner radius; 1.0, 1.25, 1.5, 1.75 for $\beta$, and 0.0005, 0.001, 0.005, 
0.01, and 0.05 \msun\ for the disk mass.  We thus consider 100 total models 
and generate SEDs at 9 inclinations\footnote{An inclination of $i=0$\degree\ 
corresponds to a pole-on (face-on) system, while an inclination of 
$i=90$\degree\ corresponds to an edge-on system.}, 
starting at $i=5$\degree\ and 
increasing in steps of 10\degree, resulting in 900 total model SEDs.  We adopt 
the dust opacities of \citet{ossenkopf1994:oh5} appropriate for thin ice 
mantles after $10^5$ yr of coagulation at a gas density of $10^6$ cm$^{-3}$ 
(OH5 dust).  The radiative transfer is performed with no external heating by 
the interstellar radiation field, but isotropic scattering off dust grains is 
included.  The input stellar spectrum is assumed to be a simple 
blackbody with $L = 1$ \lsun\ and $T_{\rm eff}$ of either 4050 K or 
3050 K \citep[corresponding approximately to K7 and M5 spectral type pre-main sequence stars, respectively;][]{pecaut2013:teff}.
Finally, for each of the 1800 total model SEDs (two stellar models each 
surrounded by 100 different circumstellar disk models, each viewed at 9 
inclinations), we calculate two values of the bolometric temperature: one with 
the full SED from optical to millimeter wavelengths, and another excluding all 
optical wavelengths (all wavelengths shortward of 1.25 \um).

\begin{figure*}
\epsscale{1.2}
\plotone{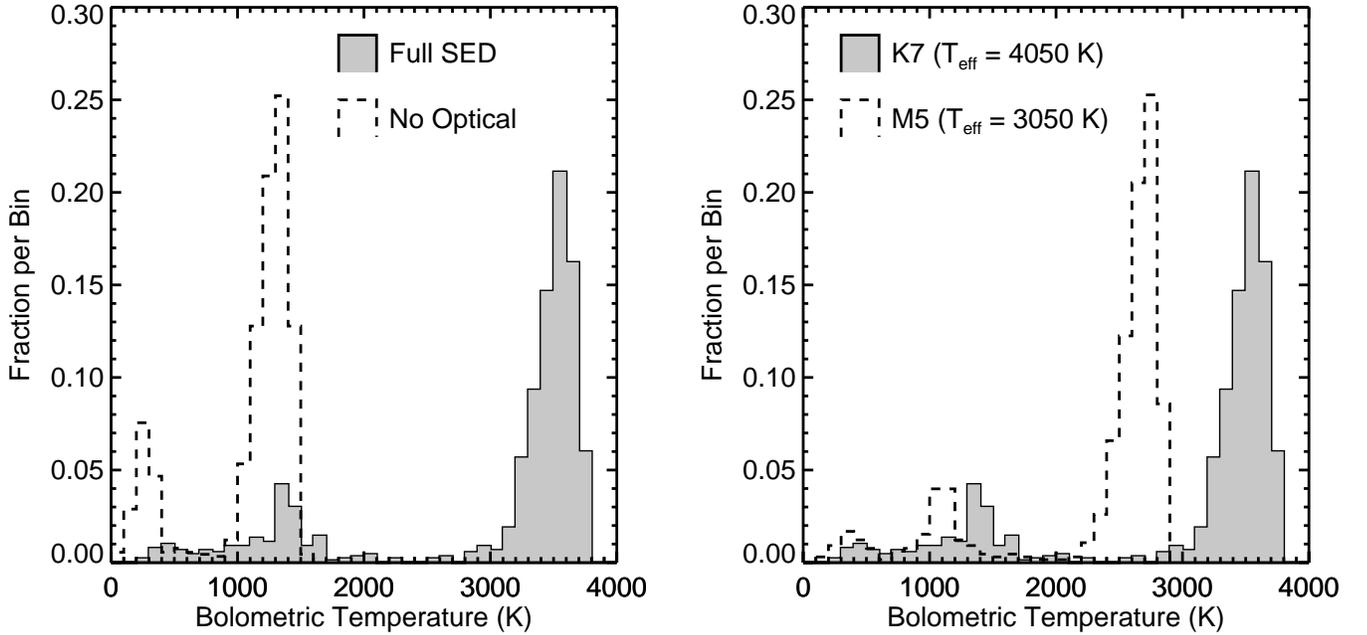}
\caption{\label{fig_tbol}Histograms showing the distribution of \tbolprime\ 
for the models described in the text.  {\it Left:} The distributions of 
\tbolprime\ for the K7 disk models, calculated both with (shaded) and without 
(dashed) optical photometry included in the SEDs.  {\it Right:} The 
distributions of \tbolprime\ for the K7 (shaded) and M5 (dashed) disk models, 
in both cases calculated with optical photometry included in the SEDs.}
\end{figure*}

The left panel of Figure \ref{fig_tbol} shows the distributions of \tbolprime\ 
for the K7 disk models, calculated both with and without optical photometry 
included in the SEDs.  Since the radiative transfer model includes no 
foreground extinction, these values are equivalent to our observed values after 
correcting for extinction.  When optical photometry is included in the SEDs, 
the distribution of \tbolprime\ exhibits a narrow peak around 3500 K with a 
width of only a few hundred K, plus a long tail to lower values corresponding 
to SEDs viewed at edge-on inclinations.  A similar behavior is exhibited 
when optical photometry is left out of the SEDs, except both the peak and long 
tail are shifted to lower values of \tbolprime.  This behavior is expected 
since the inclusion of shorter wavelength data shifts the flux-weighted mean 
frequency to higher values.

The right panel of Figure \ref{fig_tbol} shows the distributions of \tbolprime\ 
for the K7 and M5 disk models, in both cases calculated with optical 
photometry included in the SEDs.  Each set of models exhibits narrow 
peaks, with the K7 model peak located at higher values than the M5 model.  
Even for the large range of disk properties considered here, it is evident 
from this Figure that the measured \tbolprime\ for YSOs with disks is mostly 
determined by the spectral type of the central source. The disk properties 
introduce only a very narrow spread except for models viewed at nearly 
edge-on inclinations.  

The tight clustering into two narrow ranges of \tbolprime\ for most of 
our Class II and III YSOs, as evident from Figures \ref{fig_blt} and 
\ref{fig_alphatbol}, is naturally explained by these results.  A K7 
spectral type is assumed for most sources when deriving extinction 
corrections, explaining the tight clustering, and optical photometry is only 
available for a subset of our YSOs (see \S \ref{sec_method_seds}), 
explaining the bifurcation into two separate ranges of \tbolprime.  One 
consequence of this situation 
is that, even though a boundary in bolometric temperature 
between Class II and Class III is defined by \citet{chen1995:tbol}, \tbolprime\ 
is not a good discriminator between these two Classes.

\section{D.~~Tables of Ancillary Photometry Collected from the Literature}\label{sec_appendix_fluxes}

Here we tabulate all ancillary photometry used to construct YSO SEDs, where 
we define ancillary to mean all photometry beyond the 2MASS and {\it Spitzer} 
photometry measured, band-merged, and cataloged by the c2d pipeline.  This 
definition of ancillary includes {\it Spitzer} 160 \um\ photometry.  
All flux densities and flux density uncertainties are listed in mJy and 
rounded to two significant digits.  

Tables \ref{tab_fluxes_obs_optical} and \ref{tab_fluxes_dered_optical} list, 
for each YSO, the same running index as in Table \ref{tab_ysos}, followed by 
flux density and flux density uncertainty pairs for all photometry compiled 
between $0.36-0.96$ \um.  Table \ref{tab_fluxes_obs_optical} lists 
observed values whereas Table \ref{tab_fluxes_dered_optical} lists extinction 
corrected values.

Tables \ref{tab_fluxes_obs_midir} and \ref{tab_fluxes_dered_midir} list, 
for each YSO, the same running index as in Table \ref{tab_ysos}, followed by 
flux density and flux density uncertainty pairs for all ancillary photometry 
compiled between $3.4-22$ \um.  Table \ref{tab_fluxes_obs_midir} lists 
observed values whereas Table \ref{tab_fluxes_dered_midir} lists extinction 
corrected values.

Tables \ref{tab_fluxes_obs_submm} and \ref{tab_fluxes_dered_submm} list, 
for each YSO, the same running index as in Table \ref{tab_ysos}, followed by 
flux density and flux density uncertainty pairs for all ancillary photometry 
compiled between $3.4-22$ \um.  Table \ref{tab_fluxes_obs_submm} lists 
observed values whereas Table \ref{tab_fluxes_dered_submm} lists extinction 
corrected values.

\clearpage
\LongTables
\input{tab1.tex}

\input{tab2.tex}
\clearpage
\input{tab3.tex}
\input{tab4.tex}
\input{tab5.tex}

\input{tab6.tex}
\input{tab7.tex}
\input{tab8.tex}

\clearpage
\input{tab9.tex}
\input{tab10.tex}
\clearpage
\input{tab11.tex}

\input{tab12.tex}
\clearpage
\input{tab13.tex}
\input{tab14.tex}

\end{document}

%% file: tab1.tex
\begin{deluxetable*}{lcccccc}
\tabletypesize{\scriptsize}
\tablewidth{0pt}
\tablecaption{\label{tab_clouds}Molecular Clouds Surveyed by the c2d and GB Surveys}
\tablehead{
\colhead{} & \colhead{} & \colhead{Distance} & \colhead{Distance} & \colhead{Area} & \colhead{Mass} & \colhead{Data} \\
\colhead{Cloud} & \colhead{Survey} & \colhead{(pc)} & \colhead{Reference\tablenotemark{a}} & \colhead{(pc$^2$)} & \colhead{(\msun)} & \colhead{Reference(s)\tablenotemark{b}}
}
\startdata
Aquila            & GB  & 260                       & 1  & 179.0 & 24400 $\pm$ 3000 & 1 \\
Auriga/CMC        & GB  & 450                       & 2  & 52.4  & 4844 $\pm$ 1152  & 2 \\
Cepheus           & GB  & 200--325\tablenotemark{c} & 3  & 38.0  & 2610 $\pm$ 170   & 3 \\
Chamaeleon I      & GB  & 150                       & 4  & 9.4   & 857 $\pm$ 210    & \nodata\\
Chamaeleon II     & c2d & 178                       & 5  & 9.2   & 637 $\pm$ 300    & 4, 5, 6 \\
Chamaeleon III    & GB  & 150                       & 4  & 28.0  & 1330 $\pm$ 390   & \nodata\\
Corona Australis  & GB  & 130                       & 6  & 3.0   & 279 $\pm$ 110    & 7 \\
IC5146            & GB  & 950                       & 7  & 149.0 & 8550 $\pm$ 2170  & 8 \\
Lupus I           & c2d & 150                       & 8  & 8.9   & 513 $\pm$ 310    & 9, 10 \\
Lupus III         & c2d & 200                       & 8  & 15.4  & 912 $\pm$ 320    & 9, 10 \\
Lupus IV          & c2d & 150                       & 8  & 2.5   & 189 $\pm$ 95     & 9, 10 \\
Lupus V           & GB  & 150                       & 8  & 11.7  & 705 $\pm$ 220    & 11 \\
Lupus VI          & GB  & 150                       & 8  & 6.7   & 455 $\pm$ 140    & 11 \\
Musca             & GB  & 160                       & 9  & 6.8   & 335 $\pm$ 110    & \nodata \\
Ophiuchus         & c2d & 125                       & 10 & 29.6  & 3120 $\pm$ 1800  & 12 \\
Ophiuchus North   & GB  & 130                       & 11 & 7.3   & 621 $\pm$ 17     & 13 \\
Perseus           & c2d & 250                       & 12 & 73.2  & 6590 $\pm$ 3600  & 14, 15 \\
Serpens           & c2d & 429                       & 13 & 17.0  & 2340 $\pm$ 640   & 16, 17, 18
\enddata
\tablenotetext{a}{Reference for the distances quoted in this work: (1) \citet{maury2011:aquila}; (2) \citet{lada2009:california}; (3) \citet{kirk2009:cepheus}; (4) \citet{belloche2011:chami}; (5) \citet{whittet1997:dcham}; (6) \citet{neuhauser2008:cra}; (7) \citet{harvey2008:ic5146}; (8) \citet{comeron2008:lupus}; (9) \citet{knude1998:distances}; (10) \citet{wilking2008:oph}; (11) \citet{degeus1989:scocenob}; (12) \citet{enoch2006:bolocam}; (13) \citet{dzib2011:dserpens}.}
\tablenotetext{b}{References presenting the \emph{Spitzer} IRAC and MIPS observations: (1) \citet{gutermuth2008:serpsouth}; (2) \citet{broekhoven2014:auriga}; (3) \citet{kirk2009:cepheus}; (4) \citet{young2005:chamii}; (5) \citet{porras2006:chamii}; (6) \citet{alcala2008:chamii}; (7) \citet{peterson2011:cra}; (8) \citet{harvey2008:ic5146}; (9) \citet{chapman2007:lupus}; (10) \citet{merin2008:lupus}; (11) \citet{spezzi2011:lupus}; (12) \citet{padgett2008:oph}; (13) \citet{hatchell2012:ophnorth}; (14) \citet{jorgensen2006:perseus}; \citet{rebull2007:perseus}; (16) \citet{harvey2006:serpens}; (17); \citet{harvey2007:serpensmips}; (18) \citet{harvey2007:serpens}.}
\tablenotetext{c}{Different regions within Cepheus are located at different distances; see Kirk et al.~(2009) for details.}
\end{deluxetable*}

%% file: tab2.tex
\begin{deluxetable}{lccccccccccc}
\tabletypesize{\scriptsize}
\tablewidth{0pt}
\tablecaption{\label{tab_ysos}Properties of the c2d+GB YSOs}
\tablehead{
\colhead{} & \colhead{} & \colhead{Spitzer} & \colhead{} & \multicolumn{3}{c}{Observed} & \multicolumn{3}{c}{Extinction Corrected} & \colhead{} & \colhead {} \\
\colhead{} & \colhead{} & \colhead{Source Name} & \colhead{$A_{\rm V}$} & \colhead{} & \colhead{\tbol} & \colhead{\lbol} & \colhead{} & \colhead{\tbol$^{\prime}$} & \colhead{\lbol$^{\prime}$} & \colhead{Likely} & \colhead{} \\
\colhead{Index} & \colhead{Cloud} & \colhead{(SSTc2d or SSTgb +)} & \colhead{(mag)} & \colhead{$\alpha$} & \colhead{(K)} & \colhead{(\lsun)} & \colhead{$\alpha^{\prime}$} & \colhead{(K)} & \colhead{(\lsun)} & \colhead{AGB?} & \colhead{Core?}
}
\startdata
   1 &             Aquila & J180144.8-044941 &  8.2 &   -2.16 & 1700 &     3.0 &   -2.52 & 2000 &     8.2 & N & -- \\
   2 &             Aquila & J180145.8-045902 &  8.0 &   -2.14 & 1700 &     1.4 &   -2.56 & 2000 &     3.7 & N & -- \\
   3 &             Aquila & J180154.2-043753 & 12.4 &   -2.57 & 2100 &    0.52 &   -3.22 & 2600 &     4.7 & Y & -- \\
   4 &             Aquila & J180229.2-051552 &  8.3 &   -2.03 & 1700 &     3.6 &   -2.38 & 2000 &     9.8 & N & -- \\
   5 &             Aquila & J180302.0-044233 & 12.4 &   -2.45 & 2100 &    0.11 &   -3.15 & 2500 &    0.92 & N & -- \\
   6 &             Aquila & J180308.0-043523 &  8.0 &   -2.18 & 1700 &    0.24 &   -2.52 & 2000 &    0.64 & Y & -- \\
   7 &             Aquila & J180313.8-042845 &  6.9 &   -2.22 & 1800 &    0.72 &   -2.53 & 2000 &     1.7 & Y & -- \\
   8 &             Aquila & J180423.9-042248 &  9.0 &   -2.15 & 1700 &     2.2 &   -2.44 & 2000 &     6.3 & Y & -- \\
   9 &             Aquila & J180435.7-044700 &  8.1 &   -2.26 & 1700 &     1.9 &   -2.65 & 2000 &     5.3 & Y & -- \\
  10 &             Aquila & J180444.5-043706 & 12.4 &    0.53 &  540 &   0.055 &   -0.10 & 1000 &    0.11 & N & -- 
\enddata
\tablenotetext{a}{The full table is published online in a machine-readable format.  Only the first ten lines are shown here.}
\end{deluxetable}

%% file: tab3.tex
\begin{landscape}
\begin{deluxetable}{lrrrrrrrrrrrrrrrrrr}
\tabletypesize{\scriptsize}
\tablewidth{0pt}
\tablecaption{\label{tab_fluxes_obs}Observed {\it Spitzer} and 2MASS Flux Densities, in mJy, of the c2d+GB YSOs}
\tablehead{
\colhead{}      & \colhead{$J$}      & \colhead{$\Delta J$} & \colhead{$H$}      & \colhead{$\Delta H$} & \colhead{$K_s$}    & \colhead{$\Delta K_s$} & \colhead{IRAC} & \colhead{$\Delta$IRAC} & \colhead{IRAC} & \colhead{$\Delta$IRAC} & \colhead{IRAC} & \colhead{$\Delta$IRAC} & \colhead{IRAC} & \colhead{$\Delta$IRAC} & \colhead{MIPS} & \colhead{$\Delta$MIPS} & \colhead{MIPS} & \colhead{$\Delta$MIPS}\\
\colhead{Index} & \colhead{1.25 \um} & \colhead{1.25 \um}   & \colhead{1.65 \um} & \colhead{1.65 \um}    & \colhead{2.17 \um} & \colhead{2.17 \um} & \colhead{3.6 \um} & \colhead{3.6 \um} & \colhead{4.5 \um} & \colhead{4.5 \um} & \colhead{5.8 \um} & \colhead{5.8 \um} & \colhead{8.0 \um} & \colhead{8.0 \um} & \colhead{24 \um} & \colhead{24 \um} & \colhead{70 \um} & \colhead{70 \um}
}
\startdata
   1 &    290 &     7.3 &    820 &     33 &    1100 &     21 &    620 &      36 &    420 &      24 &     350 &     17 &     220 &     11 &      67 &    6.2 &     \nodata &    \nodata  \\
   2 &    140 &     3.3 &    370 &     14 &     480 &    9.3 &    300 &      15 &    160 &     7.7 &     150 &    6.9 &      92 &    4.3 &      36 &    3.3 &     \nodata &    \nodata  \\
   3 &    220 &     5.2 &    180 &    3.4 &     120 &    2.3 &     53 &     2.8 &     34 &     1.7 &      23 &    1.1 &      13 &   0.65 &     3.2 &   0.38 &       36 &     9.0  \\
   4 &    340 &     9.9 &    890 &     36 &    1200 &     30 &    800 &      41 &    530 &      26 &     490 &     24 &     330 &     16 &     100 &    9.6 &     \nodata &    \nodata  \\
   5 &     37 &     1.0 &     40 &    1.1 &      30 &   0.83 &     13 &    0.64 &    8.8 &    0.42 &     5.9 &   0.28 &     3.5 &   0.17 &     1.5 &   0.22 &     \nodata &    \nodata  \\
   6 &     23 &    0.62 &     67 &    1.6 &      81 &    1.7 &     49 &     2.5 &     33 &     1.6 &      26 &    1.3 &      19 &   0.88 &     4.7 &   0.48 &     \nodata &    \nodata  \\
   7 &     88 &     1.9 &    200 &    4.8 &     250 &    4.9 &    140 &     7.4 &     96 &     4.7 &      77 &    3.7 &      50 &    2.4 &      14 &    1.3 &     \nodata &    \nodata  \\
   8 &    180 &     4.3 &    540 &     18 &     750 &     18 &    510 &      25 &    340 &      17 &     310 &     15 &     200 &    9.5 &      48 &    4.5 &     \nodata &    \nodata  \\
   9 &    200 &     4.8 &    520 &     16 &     700 &     16 &    410 &      21 &    230 &      11 &     200 &    9.6 &     120 &    5.8 &      40 &    3.7 &     \nodata &    \nodata  \\
  10 &   0.62 &   0.050 &    2.2 &  0.090 &     3.6 &   0.14 &    4.7 &    0.23 &    6.9 &    0.33 &      12 &   0.57 &      27 &    1.3 &      67 &    6.2 &      110 &      16 
\enddata
\tablenotetext{a}{The full table is published online in a machine-readable format.  Only the first ten lines are shown here.}
\end{deluxetable}
\clearpage
\end{landscape}

%% file: tab4.tex
\begin{landscape}
\begin{deluxetable}{lrrrrrrrrrrrrrrrrrr}
\tabletypesize{\scriptsize}
\tablewidth{0pt}
\tablecaption{\label{tab_fluxes_dered}Extinction Corrected {\it Spitzer} and 2MASS Flux Densities, in mJy, of the c2d+GB YSOs}
\tablehead{
\colhead{}      & \colhead{$J$}      & \colhead{$\Delta J$} & \colhead{$H$}      & \colhead{$\Delta H$} & \colhead{$K_s$}    & \colhead{$\Delta K_s$} & \colhead{IRAC} & \colhead{$\Delta$IRAC} & \colhead{IRAC} & \colhead{$\Delta$IRAC} & \colhead{IRAC} & \colhead{$\Delta$IRAC} & \colhead{IRAC} & \colhead{$\Delta$IRAC} & \colhead{MIPS} & \colhead{$\Delta$MIPS} & \colhead{MIPS} & \colhead{$\Delta$MIPS}\\
\colhead{Index} & \colhead{1.25 \um} & \colhead{1.25 \um}   & \colhead{1.25 \um} & \colhead{1.65 \um}    & \colhead{2.17 \um} & \colhead{2.17 \um} & \colhead{3.6 \um} & \colhead{3.6 \um} & \colhead{4.5 \um} & \colhead{4.5 \um} & \colhead{5.8 \um} & \colhead{5.8 \um} & \colhead{8.0 \um} & \colhead{8.0 \um} & \colhead{24 \um} & \colhead{24 \um} & \colhead{70 \um} & \colhead{70 \um}
}
\startdata
   1 &   2200 &     55 &   2800 &    120 &    2400 &     49 &   1000 &     59 &    630 &     35 &     490 &     23 &     320 &     16 &      82 &    7.6 &     \nodata &    \nodata  \\
   2 &   1000 &     24 &   1300 &     46 &    1100 &     21 &    490 &     25 &    240 &     12 &     200 &    9.6 &     130 &    6.3 &      43 &    4.0 &     \nodata &    \nodata  \\
   3 &   4700 &    110 &   1200 &     22 &     430 &    8.3 &    110 &    5.9 &     63 &    3.1 &      39 &    1.8 &      24 &    1.2 &     4.4 &   0.51 &       38 &     9.4  \\
   4 &   2600 &     78 &   3200 &    130 &    2900 &     70 &   1300 &     68 &    800 &     40 &     680 &     34 &     490 &     23 &     130 &     12 &     \nodata &    \nodata  \\
   5 &    810 &     22 &    260 &    7.2 &     110 &    3.0 &     28 &    1.4 &     16 &   0.78 &     9.7 &   0.46 &     6.3 &   0.30 &     2.0 &   0.30 &     \nodata &    \nodata  \\
   6 &    170 &    4.5 &    220 &    5.4 &     190 &    3.9 &     79 &    4.0 &     49 &    2.3 &      37 &    1.7 &      27 &    1.3 &     5.7 &   0.58 &     \nodata &    \nodata  \\
   7 &    480 &     11 &    570 &     14 &     510 &    9.9 &    220 &     11 &    140 &    6.6 &     100 &    4.9 &      69 &    3.3 &      17 &    1.6 &     \nodata &    \nodata  \\
   8 &   1700 &     40 &   2100 &     69 &    1900 &     45 &    880 &     44 &    530 &     27 &     450 &     21 &     300 &     14 &      60 &    5.6 &     \nodata &    \nodata  \\
   9 &   1500 &     35 &   1800 &     53 &    1600 &     36 &    680 &     34 &    340 &     17 &     270 &     13 &     170 &    8.4 &      48 &    4.5 &     \nodata &    \nodata  \\
  10 &     13 &    1.1 &     14 &   0.59 &      13 &   0.50 &     10 &   0.49 &     13 &   0.61 &      20 &   0.95 &      48 &    2.2 &      90 &    8.4 &      110 &      17  
\enddata
\tablenotetext{a}{The full table is published online in a machine-readable format.  Only the first ten lines are shown here.}
\end{deluxetable}
\clearpage
\end{landscape}

%% file: tab5.tex
\begin{deluxetable*}{lcccccccccc}
\tabletypesize{\scriptsize}
\tablewidth{0pt}
\tablecaption{\label{tab_classification}Number of YSOs by Cloud and Class}
\tablehead{
\colhead{}      & \multicolumn{5}{c}{Extinction Corrected, All Sources} & \multicolumn{5}{c}{Extinction Corrected, Excluding Likely AGB Stars} \\
\colhead{Cloud} & \colhead{0+I} & \colhead{Flat} & \colhead{II} & \colhead{III} & \colhead{Total} & \colhead{0+I} & \colhead{Flat} & \colhead{II} & \colhead{III} & \colhead{Total}
}
\startdata
Aquila           & 83  & 65  & 330  & 841  & 1319 & 83  & 65  & 327  & 634  & 1109 \\
Auriga/CMC       & 35  & 8   & 73   & 17   & 133  & 35  & 8   & 73   & 15   & 131  \\
Cepheus          & 18  & 11  & 61   & 13   & 103  & 18  & 11  & 61   & 12   & 102  \\
Chamaeleon I     & 4   & 5   & 62   & 15   & 86   & 4   & 5   & 62   & 14   & 85   \\
Chamaeleon II    & 2   & 0   & 19   & 5    & 26   & 2   & 0   & 19   & 4    & 25   \\
Chamaeleon III   & 1   & 0   & 0    & 3    & 4    & 1   & 0   & 0    & 0    & 1    \\
Corona Australis & 9   & 6   & 22   & 17   & 54   & 9   & 6   & 22   & 16   & 53   \\
IC5146           & 28  & 10  & 79   & 15   & 132  & 28  & 10  & 79   & 14   & 131  \\
Lupus I          & 1   & 2   & 8    & 2    & 13   & 1   & 2   & 8    & 0    & 11   \\
Lupus III        & 2   & 5   & 38   & 24   & 69   & 2   & 5   & 38   & 14   & 59   \\
Lupus IV         & 1   & 1   & 3    & 7    & 12   & 1   & 1   & 3    & 5    & 10   \\
Lupus V          & 0   & 0   & 7    & 36   & 43   & 0   & 0   & 7    & 25   & 32   \\
Lupus VI         & 1   & 0   & 2    & 42   & 45   & 1   & 0   & 2    & 25   & 28   \\
Musca            & 1   & 0   & 1    & 11   & 13   & 1   & 0   & 1    & 4    & 6    \\
Ophiuchus        & 28  & 43  & 174  & 47   & 292  & 28  & 43  & 174  & 32   & 277  \\
Ophiuchus North  & 2   & 1   & 3    & 4    & 10   & 2   & 1   & 3    & 3    & 9    \\
Perseus          & 76  & 35  & 235  & 39   & 385  & 76  & 35  & 235  & 31   & 377  \\
Serpens          & 34  & 18  & 131  & 44   & 227  & 34  & 18  & 131  & 39   & 222  \\
Total            & 326 & 210 & 1248 & 1182 & 2966 & 326 & 210 & 1245 & 887  & 2668 
\enddata
\end{deluxetable*}

%% file: tab6.tex
\begin{deluxetable}{lcccc}
\tablewidth{0pt}
\tablecaption{\label{tab_cc2}Classification Via [3.6]$^{\prime}$~$-$~[5.8]$^{\prime}$ versus [8.0]$^{\prime}$~$-$~[24]$^{\prime}$}
\tablehead{
\colhead{Quantity}                       & \colhead{Class III/stellar} & \colhead{Class II} & \colhead{Flat} & \colhead{Class 0+I}}
\startdata
Original x$_{\rm min}$                     & $-$0.5 & 2.0    & \nodata & 3.50 \\
Original x$_{\rm max}$                     & 1.5    & 5.0    & \nodata & \nodata \\
Original y$_{\rm min}$                     & $-0.5$ & 0.3    & \nodata & 1.50 \\
Original y$_{\rm max}$                     & 0.5    & 1.5    & \nodata & \nodata \\
Original P$_{1}$\tablenotemark{a}  & 65.2\% & 77.9\% & \nodata & 66.0\% \\
Original P$_{2}$\tablenotemark{b} & 99.5\% & 84.2\% & \nodata & 84.1\% \\
Revised x$_{\rm min}$                      & $-$0.5 & 1.5    & 2.5 & 2.5 \\
Revised x$_{\rm max}$                      & 4.0    & 5.0    & 4.5 & \nodata \\
Revised y$_{\rm min}$                      & $-0.5$ & 0.4    & 1.1 & 1.9 \\
Revised y$_{\rm max}$                      & 0.4    & 1.1    & 1.8 & \nodata \\ 
Revised P$_{1}$\tablenotemark{a}          & 83.2\% & 79.2\% & 63.2\% & 65.6\% \\
Revised P$_{2}$\tablenotemark{b}          & 92.2\% & 91.8\% & 59.1\% & 90.0\%
\enddata
\tablenotetext{a}{The probability that a YSO of the desired Class is located within the region.}
\tablenotetext{b}{The probability that a YSO located within the region has the desired Class.}
\end{deluxetable} 

%% file: tab7.tex
\begin{deluxetable*}{lccccc}
\tablewidth{0pt}
\tablecaption{\label{tab_timescales}Timescales for Young Stellar Objects}
\tablehead{
\colhead{} & \colhead{$\tau_{\rm ref}$\tablenotemark{a}} & \colhead{$\tau_{\rm 0+I}$\tablenotemark{b}} & \colhead{$\tau_{\rm F}$\tablenotemark{c}} & \colhead{$\tau_{\rm 0}$\tablenotemark{d}} & \colhead{$\tau_{\rm I}$\tablenotemark{e}} \\
\colhead{Reference Class} & \colhead{(Myr)} & \colhead{(Myr)} & \colhead{(Myr)} & \colhead{(Myr)} & \colhead{(Myr)} 
}
\startdata
All Class II                     & 2.0 & 0.52 & 0.34 & 0.17 & 0.35 \\
All Class II                     & 3.0 & 0.78 & 0.50 & 0.26 & 0.52 \\
All Class II + All Class III     & 3.0 & 0.40 & 0.26 & 0.13 & 0.27 \\
All Class II + 75\% of Class III & 3.0 & 0.46 & 0.30 & 0.15 & 0.31 \\
All Class II + 10\% of Class III & 3.0 & 0.72 & 0.46 & 0.24 & 0.48 
\enddata
\tablenotetext{a}{Assumed duration of the reference Class.}
\tablenotetext{b}{Calculated duration of Class 0+I sources ($\tau_{\rm ref} \frac{N_{\rm 0+I}}{N_{\rm ref}}$).}
\tablenotetext{c}{Calculated duration of Flat-spectrum sources ($\tau_{\rm ref} \frac{N_{\rm F}}{N_{\rm ref}}$).}
\tablenotetext{d}{Calculated duration of Class 0 sources ($0.335 \times \tau_{\rm 0+I}$); see \S \ref{sec_timescales_class0} for details.}
\tablenotetext{e}{Calculated duration of Class I sources ($0.665 \times \tau_{\rm 0+I}$); see \S \ref{sec_timescales_class0} for details.}
\end{deluxetable*} 

%% file: tab8.tex
\LongTables
\begin{deluxetable}{lcccc}
\tablewidth{0pt}
\tablecaption{\label{tab_appendix_aors}\spitzer\ AORs for Unpublished Clouds}
\tablehead{
\colhead{Cloud} & \colhead{MIPS Target Name} & \colhead{MIPS AOR} & \colhead{IRAC Target Name} & \colhead{IRAC AOR}
}
\startdata
Aquila & SERAQU\_E\_1 & 19968768 & SERAQU\_E\_1 & 20008448 \\
       &              & 19968256 &              & 20008704 \\
       & SERAQU\_E\_2 & 20001536 &              & 20008960 \\
       &              & 20001024 &              & 19970816 \\
       & SERAQU\_E\_3 & 19970560 &              & 19971072 \\
       &              & 19970304 &              & 19971328 \\
       & SERAQU\_E\_4 & 20002048 & SERAQU\_E\_2 & 20005120 \\
       &              & 20001792 &              & 20005376 \\
       &              &          &              & 20005632 \\
       &              &          &              & 19964672 \\
       &              &          &              & 19965440 \\
       &              &          &              & 19965952 \\
       &              &          & SERAQU\_E\_3 & 20002304 \\
       &              &          &              & 20002560 \\
       &              &          &              & 20003072 \\
       &              &          &              & 19957248 \\
       &              &          &              & 19958016 \\
       &              &          &              & 19958272 \\
       &              &          & SERAQU\_E\_4 & 19997952 \\
       &              &          &              & 19998464 \\
       &              &          &              & 19998720 \\
       &              &          &              & 19999488 \\
       &              &          &              & 20018176 \\
       &              &          &              & 20018688 \\
       &              &          &              & 20018944 \\
       &              &          &              & 20019200 \\
       &              &          & SERAQU\_E\_5 & 14511104 \\
       &              &          & SERAQU\_E\_6 & 27042304 \\
       &              &          &              & 27042560 \\
       & SERAQU\_NE   & 14510336 & SERAQU\_NE   & 19991552 \\
       &              & 19966464 &              & 19988480 \\
       &              & 19966976 &              & 19991296 \\
       &              &          &              & 20009728 \\
       &              &          &              & 19987968 \\
       &              &          &              & 20009216 \\
       & SERAQU\_SW\_1 & 24195328 & SERAQU\_SW  & 19972352 \\
       &               & 24195584 &             & 19971840 \\
       & SERAQU\_SW\_2 & 24195840 &             & 19972608 \\
       &               & 24196096 &             & 19986944 \\
       &              &          &              & 19986688 \\
       &              &          &              & 19987200 \\
       & SERAQU\_W\_1 & 19974656 & SERAQU\_W\_1a & 19970048 \\
       &              & 19974400 &               & 19985664 \\
       &              &          & SERAQU\_W\_1b & 19962368 \\
       &              &          &               & 19980800 \\
       & SERAQU\_W\_2 & 20006912 & SERAQU\_W\_2  & 19957760 \\
       &              & 20006656 &               & 19956992 \\
       &              &          &               & 19975424 \\
       &              &          &               & 19974912 \\
Chamaeleon I & CHA\_I & 03661312 & CHA\_I\_1 & 19986432 \\
             &        & 03962112 &           & 20006400 \\
             &        & 19979264 & CHA\_I\_2 & 03960320 \\
             &        & 20011008 & CHA\_I\_3 & 20015104 \\
             &        & 19978496 &           & 20014592 \\
             &        & 20010240 & CHA\_I\_4 & 03651328 \\
             &        &          & CHA\_I\_5 & 19992832 \\
             &        &          &           & 20012800 \\
Chamaeleon III & CHA\_III\_1 & 19980032 & CHA\_III\_1 & 19995392 \\
               &             & 19979776 &             & 20015872 \\
               & CHA\_III\_2 & 20012544 & CHA\_III\_2a & 19989248 \\
               &             & 19977472 &              & 19989760 \\
               &             & 20012032 &              & 20007168 \\
               &             & 19976448 &              & 20007424 \\
               &             &          & CHA\_III\_2b & 19985152 \\
               &             &          &              & 20003328 \\
               &             &          & CHA\_III\_2c & 19961344 \\
               &             &          &              & 19977216 \\
               &             &          &              & 19984896 \\
               &             &          &              & 19961600 \\
               &             &          &              & 19998208 \\
               &             &          &              & 20002816 \\
               &             &          &              & 27040256 \\
               &             &          &              & 27040512 \\
               &             &          &              & 27041536 \\
               &             &          &              & 27042048 \\
Musca          & MUSCA & 20004096 & MUSCA\_1 & 27040768 \\
               &       & 19973120 &          & 27041024 \\
               &       & 20003840 & MUSCA\_2 & 27043072 \\
               &       & 19972864 &          & 27043328 \\
               &       &          & MUSCA\_3 & 20009472 \\
               &       &          &          & 19975680 \\
               &       &          &          & 19976704 \\
               &       &          &          & 20009984 \\
               &       &          &          & 19975936 \\
               &       &          &          & 19977728
\enddata
\end{deluxetable}

%% file: tab9.tex
\begin{landscape}
\begin{deluxetable}{lrrrrrrrrrrrrrrrr}
\tabletypesize{\scriptsize}
\tablewidth{0pt}
\tablecaption{\label{tab_fluxes_obs_optical}Observed $0.36-0.96$ \um\ Flux Densities, in mJy, of the c2d+GB YSOs}
\tablehead{
\colhead{}      & \colhead{$U$}      & \colhead{$\Delta U$} & \colhead{$B$}      & \colhead{$\Delta B$} & \colhead{$V$}    & \colhead{$\Delta V$} & \colhead{$R$} & \colhead{$\Delta R$} & \colhead{$H\alpha 12$} & \colhead{$\Delta H\alpha 12$} & \colhead{$I$} & \colhead{$\Delta I$} & \colhead{$m914$} & \colhead{$\Delta m914$} & \colhead{$z$} & \colhead{$\Delta z$}\\
\colhead{Index} & \colhead{0.36 \um} & \colhead{0.36 \um}   & \colhead{0.44 \um} & \colhead{0.44 \um}    & \colhead{0.55 \um} & \colhead{0.55 \um} & \colhead{0.64 \um} & \colhead{0.64 \um} & \colhead{0.665 \um} & \colhead{0.665 \um} & \colhead{0.79 \um} & \colhead{0.79 \um} & \colhead{0.915 \um} & \colhead{0.915 \um} & \colhead{0.96 \um} & \colhead{0.96 \um}
}
\startdata
   1 & \nodata &   \nodata &   \nodata &    \nodata &   \nodata &   \nodata &   \nodata &   \nodata &   \nodata &   \nodata &  \nodata &   \nodata &  \nodata &   \nodata &  \nodata &   \nodata  \\
   2 & \nodata &   \nodata &   \nodata &    \nodata &   \nodata &   \nodata &   \nodata &   \nodata &   \nodata &   \nodata &  \nodata &   \nodata &  \nodata &   \nodata &  \nodata &   \nodata  \\
   3 & \nodata &   \nodata &   \nodata &    \nodata &   \nodata &   \nodata &   \nodata &   \nodata &   \nodata &   \nodata &  \nodata &   \nodata &  \nodata &   \nodata &  \nodata &   \nodata  \\
   4 & \nodata &   \nodata &   \nodata &    \nodata &   \nodata &   \nodata &   \nodata &   \nodata &   \nodata &   \nodata &  \nodata &   \nodata &  \nodata &   \nodata &  \nodata &   \nodata  \\
   5 & \nodata &   \nodata &   \nodata &    \nodata &   \nodata &   \nodata &   \nodata &   \nodata &   \nodata &   \nodata &  \nodata &   \nodata &  \nodata &   \nodata &  \nodata &   \nodata  \\
   6 & \nodata &   \nodata &   \nodata &    \nodata &   \nodata &   \nodata &   \nodata &   \nodata &   \nodata &   \nodata &  \nodata &   \nodata &  \nodata &   \nodata &  \nodata &   \nodata  \\
   7 & \nodata &   \nodata &   \nodata &    \nodata &   \nodata &   \nodata &   \nodata &   \nodata &   \nodata &   \nodata &  \nodata &   \nodata &  \nodata &   \nodata &  \nodata &   \nodata  \\
   8 & \nodata &   \nodata &   \nodata &    \nodata &   \nodata &   \nodata &   \nodata &   \nodata &   \nodata &   \nodata &  \nodata &   \nodata &  \nodata &   \nodata &  \nodata &   \nodata  \\
   9 & \nodata &   \nodata &   \nodata &    \nodata &   \nodata &   \nodata &   \nodata &   \nodata &   \nodata &   \nodata &  \nodata &   \nodata &  \nodata &   \nodata &  \nodata &   \nodata  \\
  10 & \nodata &   \nodata &   \nodata &    \nodata &   \nodata &   \nodata &   \nodata &   \nodata &   \nodata &   \nodata &  \nodata &   \nodata &  \nodata &   \nodata &  \nodata &   \nodata 
\enddata
\tablenotetext{a}{The full table is published online in a machine-readable format.  Only the first ten lines are shown here.}
\end{deluxetable}
\clearpage
\end{landscape}

%% file: tab10.tex
\begin{landscape}
\begin{deluxetable}{lrrrrrrrrrrrrrrrr}
\tabletypesize{\scriptsize}
\tablewidth{0pt}
\tablecaption{\label{tab_fluxes_dered_optical}Extinction Corrected $0.36-0.96$ \um\ Flux Densities, in mJy, of the c2d+GB YSOs}
\tablehead{
\colhead{}      & \colhead{$U$}      & \colhead{$\Delta U$} & \colhead{$B$}      & \colhead{$\Delta B$} & \colhead{$V$}    & \colhead{$\Delta V$} & \colhead{$R$} & \colhead{$\Delta R$} & \colhead{$H\alpha 12$} & \colhead{$\Delta H\alpha 12$} & \colhead{$I$} & \colhead{$\Delta I$} & \colhead{$m914$} & \colhead{$\Delta m914$} & \colhead{$z$} & \colhead{$\Delta z$}\\
\colhead{Index} & \colhead{0.36 \um} & \colhead{0.36 \um}   & \colhead{0.44 \um} & \colhead{0.44 \um}    & \colhead{0.55 \um} & \colhead{0.55 \um} & \colhead{0.64 \um} & \colhead{0.64 \um} & \colhead{0.665 \um} & \colhead{0.665 \um} & \colhead{0.79 \um} & \colhead{0.79 \um} & \colhead{0.915 \um} & \colhead{0.915 \um} & \colhead{0.96 \um} & \colhead{0.96 \um}
}
\startdata
   1 & \nodata &   \nodata &   \nodata &   \nodata &  \nodata &  \nodata &  \nodata &   \nodata &  \nodata &  \nodata & \nodata &   \nodata & \nodata &   \nodata &  \nodata &   \nodata  \\
   2 & \nodata &   \nodata &   \nodata &   \nodata &  \nodata &  \nodata &  \nodata &   \nodata &  \nodata &  \nodata & \nodata &   \nodata & \nodata &   \nodata &  \nodata &   \nodata  \\
   3 & \nodata &   \nodata &   \nodata &   \nodata &  \nodata &  \nodata &  \nodata &   \nodata &  \nodata &  \nodata & \nodata &   \nodata & \nodata &   \nodata &  \nodata &   \nodata  \\
   4 & \nodata &   \nodata &   \nodata &   \nodata &  \nodata &  \nodata &  \nodata &   \nodata &  \nodata &  \nodata & \nodata &   \nodata & \nodata &   \nodata &  \nodata &   \nodata  \\
   5 & \nodata &   \nodata &   \nodata &   \nodata &  \nodata &  \nodata &  \nodata &   \nodata &  \nodata &  \nodata & \nodata &   \nodata & \nodata &   \nodata &  \nodata &   \nodata  \\
   6 & \nodata &   \nodata &   \nodata &   \nodata &  \nodata &  \nodata &  \nodata &   \nodata &  \nodata &  \nodata & \nodata &   \nodata & \nodata &   \nodata &  \nodata &   \nodata  \\
   7 & \nodata &   \nodata &   \nodata &   \nodata &  \nodata &  \nodata &  \nodata &   \nodata &  \nodata &  \nodata & \nodata &   \nodata & \nodata &   \nodata &  \nodata &   \nodata  \\
   8 & \nodata &   \nodata &   \nodata &   \nodata &  \nodata &  \nodata &  \nodata &   \nodata &  \nodata &  \nodata & \nodata &   \nodata & \nodata &   \nodata &  \nodata &   \nodata  \\
   9 & \nodata &   \nodata &   \nodata &   \nodata &  \nodata &  \nodata &  \nodata &   \nodata &  \nodata &  \nodata & \nodata &   \nodata & \nodata &   \nodata &  \nodata &   \nodata  \\
  10 & \nodata &   \nodata &   \nodata &   \nodata &  \nodata &  \nodata &  \nodata &   \nodata &  \nodata &  \nodata & \nodata &   \nodata & \nodata &   \nodata &  \nodata &   \nodata 
\enddata
\tablenotetext{a}{The full table is published online in a machine-readable format.  Only the first ten lines are shown here.}
\end{deluxetable}
\clearpage
\end{landscape}

%% file: tab11.tex
\begin{deluxetable}{lrrrrrrrrrrrr}
\tabletypesize{\scriptsize}
\tablewidth{0pt}
\tablecaption{\label{tab_fluxes_obs_midir}Observed $3.4-22$ \um\ Flux Densities, in mJy, of the c2d+GB YSOs}
\tablehead{
\colhead{}      & \colhead{$L$}      & \colhead{$\Delta L$} & \colhead{$M$}      & \colhead{$\Delta M$} & \colhead{$ISO$}    & \colhead{$\Delta ISO$} & \colhead{$WISE$} & \colhead{$\Delta WISE$} & \colhead{$ISO$} & \colhead{$\Delta ISO$} & \colhead{$WISE$} & \colhead{$\Delta WISE$}\\
\colhead{Index} & \colhead{3.4 \um} & \colhead{3.4 \um}   & \colhead{5.0 \um} & \colhead{5.0 \um}    & \colhead{6.7 \um} & \colhead{6.7 \um} & \colhead{12 \um} & \colhead{12 \um} & \colhead{14 \um} & \colhead{14 \um} & \colhead{22 \um} & \colhead{22 \um}
}
\startdata
   1 &  \nodata &  \nodata &  \nodata &    \nodata & \nodata & \nodata &    160 &   2.3 & \nodata &   \nodata &     87 &  2.3  \\
   2 &  \nodata &  \nodata &  \nodata &    \nodata & \nodata & \nodata &     70 &   1.0 & \nodata &   \nodata &     47 &  1.6  \\
   3 &  \nodata &  \nodata &  \nodata &    \nodata & \nodata & \nodata &    7.6 &  0.21 & \nodata &   \nodata &    3.7 & 0.99  \\
   4 &  \nodata &  \nodata &  \nodata &    \nodata & \nodata & \nodata &    240 &   3.3 & \nodata &   \nodata &    120 &  2.9  \\
   5 &  \nodata &  \nodata &  \nodata &    \nodata & \nodata & \nodata &    1.7 &  0.16 & \nodata &   \nodata &   \nodata & \nodata  \\
   6 &  \nodata &  \nodata &  \nodata &    \nodata & \nodata & \nodata &     14 &  0.28 & \nodata &   \nodata &    8.2 & 0.85  \\
   7 &  \nodata &  \nodata &  \nodata &    \nodata & \nodata & \nodata &     33 &  0.51 & \nodata &   \nodata &     19 &  1.1  \\
   8 &  \nodata &  \nodata &  \nodata &    \nodata & \nodata & \nodata &    120 &   1.5 & \nodata &   \nodata &     63 &  2.1  \\
   9 &  \nodata &  \nodata &  \nodata &    \nodata & \nodata & \nodata &     81 &   1.2 & \nodata &   \nodata &     50 &  1.7  \\
  10 &  \nodata &  \nodata &  \nodata &    \nodata & \nodata & \nodata &     27 &  0.47 & \nodata &   \nodata &     65 &  2.2 
\enddata
\tablenotetext{a}{The full table is published online in a machine-readable format.  Only the first ten lines are shown here.}
\end{deluxetable}

%% file: tab12.tex
%\begin{landscape}
\begin{deluxetable}{lrrrrrrrrrrrr}
\tabletypesize{\scriptsize}
\tablewidth{0pt}
\tablecaption{\label{tab_fluxes_dered_midir}Extinction Corrected $3.4-22$ \um\ Flux Densities, in mJy, of the c2d+GB YSOs}
\tablehead{
\colhead{}      & \colhead{$L$}      & \colhead{$\Delta L$} & \colhead{$M$}      & \colhead{$\Delta M$} & \colhead{$ISO$}    & \colhead{$\Delta ISO$} & \colhead{$WISE$} & \colhead{$\Delta WISE$} & \colhead{$ISO$} & \colhead{$\Delta ISO$} & \colhead{$WISE$} & \colhead{$\Delta WISE$}\\
\colhead{Index} & \colhead{3.4 \um} & \colhead{3.4 \um}   & \colhead{5.0 \um} & \colhead{5.0 \um}    & \colhead{6.7 \um} & \colhead{6.7 \um} & \colhead{12 \um} & \colhead{12 \um} & \colhead{14 \um} & \colhead{14 \um} & \colhead{22 \um} & \colhead{22 \um}
}
\startdata
   1 &  \nodata &  \nodata &  \nodata &    \nodata & \nodata & \nodata &    230 &   3.4 & \nodata &   \nodata &    110 &  2.9  \\
   2 &  \nodata &  \nodata &  \nodata &    \nodata & \nodata & \nodata &    100 &   1.5 & \nodata &   \nodata &     59 &  2.0  \\
   3 &  \nodata &  \nodata &  \nodata &    \nodata & \nodata & \nodata &     14 &  0.39 & \nodata &   \nodata &    5.2 &  1.4  \\
   4 &  \nodata &  \nodata &  \nodata &    \nodata & \nodata & \nodata &    360 &   4.9 & \nodata &   \nodata &    160 &  3.6  \\
   5 &  \nodata &  \nodata &  \nodata &    \nodata & \nodata & \nodata &    3.1 &  0.29 & \nodata &   \nodata &   \nodata & \nodata  \\
   6 &  \nodata &  \nodata &  \nodata &    \nodata & \nodata & \nodata &     21 &  0.42 & \nodata &   \nodata &     10 &  1.1  \\
   7 &  \nodata &  \nodata &  \nodata &    \nodata & \nodata & \nodata &     46 &  0.72 & \nodata &   \nodata &     23 &  1.3  \\
   8 &  \nodata &  \nodata &  \nodata &    \nodata & \nodata & \nodata &    190 &   2.4 & \nodata &   \nodata &     81 &  2.7  \\
   9 &  \nodata &  \nodata &  \nodata &    \nodata & \nodata & \nodata &    120 &   1.8 & \nodata &   \nodata &     62 &  2.1  \\
  10 &  \nodata &  \nodata &  \nodata &    \nodata & \nodata & \nodata &     49 &  0.87 & \nodata &   \nodata &     92 &  3.2  
\enddata
\tablenotetext{a}{The full table is published online in a machine-readable format.  Only the first ten lines are shown here.}
\end{deluxetable}
%\clearpage
%\end{landscape}

%% file: tab13.tex
\begin{landscape}
\begin{deluxetable}{lrrrrrrrrrrrrrr}
\tabletypesize{\scriptsize}
\tablewidth{0pt}
\tablecaption{\label{tab_fluxes_obs_submm}Observed $160-1300$ \um\ Flux Densities, in mJy, of the c2d+GB YSOs}
\tablehead{
\colhead{}      & \colhead{$MIPS$}      & \colhead{$\Delta MIPS$} & \colhead{}      & \colhead{} & \colhead{}    & \colhead{} & \colhead{} & \colhead{} & \colhead{} & \colhead{} & \colhead{} & \colhead{} & \colhead{} & \colhead{} \\
\colhead{Index} & \colhead{160 \um} & \colhead{160 \um}   & \colhead{350 \um} & \colhead{$\Delta$350 \um}    & \colhead{450 \um} & \colhead{$\Delta$450 \um} & \colhead{850 \um} & \colhead{$\Delta$850 \um} & \colhead{1100 \um} & \colhead{$\Delta$1100 \um} & \colhead{1200 \um} & \colhead{$\Delta$1200 \um} & \colhead{1300 \um} & \colhead{$\Delta$1300 \um}
}
\startdata
   1 &  \nodata &  \nodata &   \nodata &  \nodata &   \nodata &  \nodata &  \nodata & \nodata & \nodata & \nodata &   \nodata & \nodata &   \nodata & \nodata  \\
   2 &  \nodata &  \nodata &   \nodata &  \nodata &   \nodata &  \nodata &  \nodata & \nodata & \nodata & \nodata &   \nodata & \nodata &   \nodata & \nodata  \\
   3 &  \nodata &  \nodata &   \nodata &  \nodata &   \nodata &  \nodata &  \nodata & \nodata & \nodata & \nodata &   \nodata & \nodata &   \nodata & \nodata  \\
   4 &  \nodata &  \nodata &   \nodata &  \nodata &   \nodata &  \nodata &  \nodata & \nodata & \nodata & \nodata &   \nodata & \nodata &   \nodata & \nodata  \\
   5 &  \nodata &  \nodata &   \nodata &  \nodata &   \nodata &  \nodata &  \nodata & \nodata & \nodata & \nodata &   \nodata & \nodata &   \nodata & \nodata  \\
   6 &  \nodata &  \nodata &   \nodata &  \nodata &   \nodata &  \nodata &  \nodata & \nodata & \nodata & \nodata &   \nodata & \nodata &   \nodata & \nodata  \\
   7 &  \nodata &  \nodata &   \nodata &  \nodata &   \nodata &  \nodata &  \nodata & \nodata & \nodata & \nodata &   \nodata & \nodata &   \nodata & \nodata  \\
   8 &  \nodata &  \nodata &   \nodata &  \nodata &   \nodata &  \nodata &  \nodata & \nodata & \nodata & \nodata &   \nodata & \nodata &   \nodata & \nodata  \\
   9 &  \nodata &  \nodata &   \nodata &  \nodata &   \nodata &  \nodata &  \nodata & \nodata & \nodata & \nodata &   \nodata & \nodata &   \nodata & \nodata  \\
  10 &  \nodata &  \nodata &   \nodata &  \nodata &   \nodata &  \nodata &  \nodata & \nodata & \nodata & \nodata &   \nodata & \nodata &   \nodata & \nodata  
\enddata
\tablenotetext{a}{The full table is published online in a machine-readable format.  Only the first ten lines are shown here.}
\end{deluxetable}
\clearpage
\end{landscape}

%% file: tab14.tex
\begin{landscape}
\begin{deluxetable}{lrrrrrrrrrrrrrr}
\tabletypesize{\scriptsize}
\tablewidth{0pt}
\tablecaption{\label{tab_fluxes_dered_submm}Extinction Corrected $160-1300$ \um\ Flux Densities, in mJy, of the c2d+GB YSOs}
\tablehead{
\colhead{}      & \colhead{$MIPS$}      & \colhead{$\Delta MIPS$} & \colhead{}      & \colhead{} & \colhead{}    & \colhead{} & \colhead{} & \colhead{} & \colhead{} & \colhead{} & \colhead{} & \colhead{} & \colhead{} & \colhead{} \\
\colhead{Index} & \colhead{160 \um} & \colhead{160 \um}   & \colhead{350 \um} & \colhead{$\Delta$350 \um}    & \colhead{450 \um} & \colhead{$\Delta$450 \um} & \colhead{850 \um} & \colhead{$\Delta$850 \um} & \colhead{1100 \um} & \colhead{$\Delta$1100 \um} & \colhead{1200 \um} & \colhead{$\Delta$1200 \um} & \colhead{1300 \um} & \colhead{$\Delta$1300 \um}
}
\startdata
   1 &  \nodata &  \nodata &   \nodata &  \nodata &   \nodata &  \nodata &  \nodata & \nodata & \nodata & \nodata &   \nodata & \nodata &   \nodata & \nodata  \\
   2 &  \nodata &  \nodata &   \nodata &  \nodata &   \nodata &  \nodata &  \nodata & \nodata & \nodata & \nodata &   \nodata & \nodata &   \nodata & \nodata  \\
   3 &  \nodata &  \nodata &   \nodata &  \nodata &   \nodata &  \nodata &  \nodata & \nodata & \nodata & \nodata &   \nodata & \nodata &   \nodata & \nodata  \\
   4 &  \nodata &  \nodata &   \nodata &  \nodata &   \nodata &  \nodata &  \nodata & \nodata & \nodata & \nodata &   \nodata & \nodata &   \nodata & \nodata  \\
   5 &  \nodata &  \nodata &   \nodata &  \nodata &   \nodata &  \nodata &  \nodata & \nodata & \nodata & \nodata &   \nodata & \nodata &   \nodata & \nodata  \\
   6 &  \nodata &  \nodata &   \nodata &  \nodata &   \nodata &  \nodata &  \nodata & \nodata & \nodata & \nodata &   \nodata & \nodata &   \nodata & \nodata  \\
   7 &  \nodata &  \nodata &   \nodata &  \nodata &   \nodata &  \nodata &  \nodata & \nodata & \nodata & \nodata &   \nodata & \nodata &   \nodata & \nodata  \\
   8 &  \nodata &  \nodata &   \nodata &  \nodata &   \nodata &  \nodata &  \nodata & \nodata & \nodata & \nodata &   \nodata & \nodata &   \nodata & \nodata  \\
   9 &  \nodata &  \nodata &   \nodata &  \nodata &   \nodata &  \nodata &  \nodata & \nodata & \nodata & \nodata &   \nodata & \nodata &   \nodata & \nodata  \\
  10 &  \nodata &  \nodata &   \nodata &  \nodata &   \nodata &  \nodata &  \nodata & \nodata & \nodata & \nodata &   \nodata & \nodata &   \nodata & \nodata  
\enddata
\tablenotetext{a}{The full table is published online in a machine-readable format.  Only the first ten lines are shown here.}
\end{deluxetable}
\clearpage
\end{landscape}